\documentclass[floatfix,
 aps, prc,     
 showpacs, 
 showkeys, 
 superscriptaddress, 
 reprint, 
 nofootinbib 
]{revtex4-1}

\usepackage[english]{babel}
\usepackage[T1]{fontenc}
\usepackage[utf8]{inputenc} 
\usepackage{amsmath} 
\usepackage{amssymb, bbm} 
\usepackage{amstext} 
\usepackage{mathrsfs} 
\usepackage{booktabs} 
\usepackage{array} 
\usepackage{graphicx} \graphicspath{ {./files/} } 
\usepackage{xcolor} \definecolor{lblue}{RGB}{70,126,185}
\usepackage[pdftex,pdfpagelabels]{hyperref} 
\hypersetup{
   colorlinks=true, linktocpage=true, urlcolor=lblue, linkcolor=lblue, citecolor=lblue,
   pdftitle={Leading three-baryon forces from SU(3) chiral effective field theory},
   pdfauthor={Stefan Petschauer}
}
\usepackage{afterpage}
\usepackage{overpic}

\newcommand{\Ps}{P^{(\sigma)}}
\newcommand{\Pt}{P^{(\tau)}}
\newcommand{\Pp}{P^{(p)}}
\newcommand{\bBa}{\bar B_\alpha}
\newcommand{\bBb}{\bar B_\beta}
\newcommand{\bBg}{\bar B_\gamma}
\newcommand{\Ba}{(\Gamma^{1,a} B)_\alpha}
\newcommand{\Bb}{(\Gamma^{2,a} B)_\beta}
\newcommand{\Bg}{(\Gamma^{3,a} B)_\gamma}
\newcommand{\ir}[1]{\mathbf{#1}}
\newcommand{\extfield}{(\nabla^i \phi)}
\newcommand{\BBBOMEpic}[6]{
  \parbox[c][2.35cm][c]{1.7cm}{
  \begin{overpic}[scale=.37]{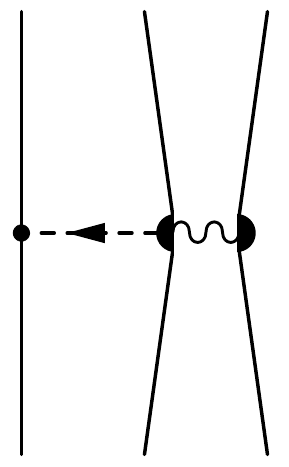}
    \put(-10,104){$#4$}\put(20,104){$#5$}\put(50,104){$#6$}
    \put(-10,-15){$#1$}\put(20,-15){$#2$}\put(50,-15){$#3$}
  \end{overpic}}
}
\newcommand{\BBBTMEpic}[6]{
  \parbox[c][2.35cm][c]{1.7cm}{
  \begin{overpic}[scale=.37]{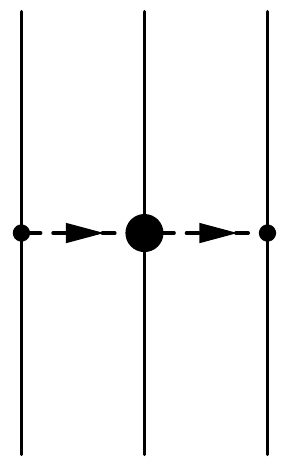}
    \put(-10,104){$#4$}\put(20,104){$#5$}\put(50,104){$#6$}
    \put(-10,-15){$#1$}\put(20,-15){$#2$}\put(50,-15){$#3$}
  \end{overpic}}
}
\newcommand{\trace}[1]{\ensuremath{\langle #1\rangle}}
\newcommand\numberthis{\addtocounter{equation}{1}\tag{\theequation}}
\makeatletter
\protected\def\lc{C}
\protected\def\ld{D}
\newcommand*{\balancecolsandclearpage}{%
  \close@column@grid
  \cleardoublepage
  \twocolumngrid
}
\protected\def\fig{\@ifstar\@fig\@@fig} \def\@fig#1{\ref{#1}} \def\@@fig#1{Fig.~\ref{#1}}
\protected\def\tab{\@ifstar\@tab\@@tab} \def\@tab#1{\ref{#1}} \def\@@tab#1{Tab.~\ref{#1}}
\protected\def\eq{\@ifstar\@eq\@@eq} \def\@eq#1{\eqref{#1}} \def\@@eq#1{Eq.~\eqref{#1}}
\protected\def\ch{\@ifstar\@ch\@@ch} \def\@ch#1{\ref{#1}} \def\@@ch#1{Ch.~\ref{#1}}
\protected\def\sect{\@ifstar\@sect\@@sect} \def\@sect#1{\ref{#1}} \def\@@sect#1{Sec.~\ref{#1}}
\protected\def\ssect{\@ifstar\@ssect\@@ssect} \def\@ssect#1{\ref{#1}} \def\@@ssect#1{Subsec.~\ref{#1}}
\protected\def\app{\@ifstar\@app\@@app} \def\@app#1{\cite{#1}} \def\@@app#1{App.~\ref{#1}}
\protected\def\ct{\@ifstar\@ct\@@ct} \def\@ct#1{\cite{#1}} \def\@@ct#1{Ref.~\cite{#1}}
\protected\def\figs{\@ifstar\@figs\@@figs} \def\@figs#1{\ref{#1}} \def\@@figs#1{Figs.~\ref{#1}}
\protected\def\tabs{\@ifstar\@tabs\@@tabs} \def\@tabs#1{\ref{#1}} \def\@@tabs#1{Tabs.~\ref{#1}}
\protected\def\eqs{\@ifstar\@eqs\@@eqs} \def\@eqs#1{\eqref{#1}} \def\@@eqs#1{Eqs.~\eqref{#1}}
\protected\def\chs{\@ifstar\@chs\@@chs} \def\@chs#1{\ref{#1}} \def\@@chs#1{Chs.~\ref{#1}}
\protected\def\sects{\@ifstar\@sects\@@sects} \def\@sects#1{\ref{#1}} \def\@@sects#1{Secs.~\ref{#1}}
\protected\def\ssects{\@ifstar\@ssects\@@ssects} \def\@ssects#1{\ref{#1}} \def\@@ssects#1{Subsecs.~\ref{#1}}
\protected\def\apps{\@ifstar\@apps\@@apps} \def\@apps#1{\cite{#1}} \def\@@apps#1{Apps.~\ref{#1}}
\protected\def\cts{\@ifstar\@cts\@@cts} \def\@cts#1{\cite{#1}} \def\@@cts#1{Refs.~\cite{#1}}
\makeatother

\newcommand{\Tab}[1]{Table~\ref{#1}}

\newcommand{\Sect}[1]{Section~\ref{#1}}

\newcommand{\ie}{i.e., }

\newcommand{\eg}{e.g., }

\newcommand{\cf}{cf.\ }

\newcommand{\cheft}{\(\chi\)EFT}

\begin{document}

\title{Leading three-baryon forces from SU(3) chiral effective field theory}

\author{Stefan~Petschauer}
   \email{stefan.petschauer@ph.tum.de}
   \affiliation{Physik Department, Technische Universit\"at M\"unchen, D-85747 Garching, Germany}
\author{Norbert~Kaiser}
   \affiliation{Physik Department, Technische Universit\"at M\"unchen, D-85747 Garching, Germany}
\author{Johann~Haidenbauer}
   \affiliation{Institute for Advanced Simulation and J\"ulich Center for Hadron Physics,\\
                Institut f\"ur Kernphysik, Forschungszentrum J\"ulich, D-52425 J\"ulich, Germany}
\author{Ulf-G.~Mei\ss{}ner}
   \affiliation{Institute for Advanced Simulation and J\"ulich Center for Hadron Physics,\\
                Institut f\"ur Kernphysik, Forschungszentrum J\"ulich, D-52425 J\"ulich, Germany}
   \affiliation{Helmholtz-Institut f\"ur Strahlen- und Kernphysik and Bethe Center\\
                for Theoretical Physics, Universit\"at Bonn, D-53115 Bonn, Germany}
\author{Wolfram~Weise}
   \affiliation{Physik Department, Technische Universit\"at M\"unchen, D-85747 Garching, Germany}
   \affiliation{ECT*, Villa Tambosi, 38123 Villazzano (Trento), Italy}

\date{\today}

\begin{abstract}
Leading three-baryon forces are derived within SU(3) chiral effective field theory.
Three classes of irreducible diagrams contribute: three-baryon contact terms, one-meson exchange and two-meson exchange diagrams.
We provide the minimal non-relativistic terms of the chiral Lagrangian, that contribute to these diagrams.
SU(3) relations are given for the strangeness $S=0$ and $-1$ sectors.
In the strangeness-zero sector we recover the well-known three-nucleon forces from chiral effective field theory.
Explicit expressions for the $\Lambda NN$ chiral potential in isospin space are presented.
\end{abstract}

\pacs{
12.39.Fe 
13.75.Ev 
14.20.Jn 
21.30.-x 
}

\keywords{chiral effective field theory, three-baryon forces, hyperons}

\maketitle

\section{Introduction}

Solving nuclear few- and many-body problems based on microscopic interactions has been a continuous challenge in nuclear physics.
Nowadays the nucleon-nucleon (\(NN\)) interaction can be treated to high accuracy using phenomenological models \ct*{Stoks1994,Wiringa1995,Machleidt2000} or potentials derived from chiral effective field theory (\cheft) \ct*{Epelbaum2009,Machleidt2011}.
However, few-body systems such as the triton cannot be described satisfactorily with two-body forces only.
Substantial improvements result from the consideration of three-nucleon forces (3NF) \ct*{Pieper2001a,Epelbaum2002}.
These 3NF are introduced either phenomenologically, such as the families of Tuscon-Melbourne \ct*{McKellar1968,Coon1975}, Brazilian \ct*{Coelho1983} or Urbana-Illinois \ct*{Pudliner1997,Pieper2001} 3NFs, or deduced from more basic principles using \cheft{} \ct*{Weinberg1990,Weinberg1991,Weinberg1992,VanKolck1994,Epelbaum2002,Ishikawa2007,Bernard2008,Bernard2011,Krebs2012,Krebs2013a}.
Effective field theory approaches have the advantage that three-nucleon forces can be derived consistently with the underlying \(NN\) interaction, and that theoretical error estimates are possible.

The situation in strangeness nuclear physics is less clear.
Due to the lack of high-precision experimental data, the hyperon-nucleon (\(YN\)) interaction cannot be sufficiently well constrained.
Different models describe the empirical scattering data equivalently \ct*{Kohno1999,Haidenbauer2005,Rijken2010,Haidenbauer2013a}, but differ considerably from each other.
Nonetheless three-baryon forces (3BF), in particular a repulsive \(\Lambda NN\) interaction, appear to be essential for the description of hypernuclei and hypernuclear matter
\ct*{Bhaduri1967,Bhaduri1967a,Gal1971,Gal1972,Gal1978,Bodmer1988,Usmani1995,Lonardoni2013,Lonardoni2013c}.
Empirical facts about dense neutron star matter favor such considerations. The recent observation of two-solar-mass neutron stars \ct*{Demorest2010,Antoniadis2013} sets strong stiffness constraints for the equation-of-state (EoS) of dense baryonic matter \ct*{Hebeler2010b, Hell2014, *[{}] [{ and refs.\ therein.}]  Steiner2015}.
A naive introduction of \(\Lambda\)-hyperons as an additional baryonic degree of freedom in neutron star matter would soften the EoS \ct*{Djapo2010} such that it is not possible to stabilize two-solar-mass neutron stars against gravitational collapse.
The introduction of strongly repulsive hyperon-nucleon-nucleon forces is one possible suggestion to improve the situation \ct*{Takatsuka2008,Vidana2010,Lonardoni2015}.

So far, baryonic three-body forces involving hyperons have been investigated only by employing phenomenological interactions, and a more systematic approach is desirable. Chiral effective field theory is an appropriate tool for such considerations. 
It exploits the symmetries of quantum chromodynamics together with the appropriate low-energy degrees of freedom.
The description of the low-energy interaction of hadrons can be improved systematically by going to higher order in the power counting in small momenta.
Furthermore, the hierarchy of baryonic forces, from long-range to intermediate- and short-range interactions, emerges naturally within this framework. Two- and three-baryon forces can be described in a consistent way.

Recently, the hyperon-nucleon interaction has been studied up to next-to-leading order (NLO) in \cheft.
The \(YN\) scattering data \ct*{Haidenbauer2013a}, as well as the self-energies of hyperons in nuclear matter \ct*{Haidenbauer2015a,Petschauer2015}, can be well described within this framework.
The irreducible chiral three-baryon forces appear formally at next-to-next-to-leading order (NNLO) \ct*{Epelbaum2009}.
However, \eg the low-energy constants of the three-nucleon forces at NNLO are unnaturally large
and cause effects comparable to those one would expect at the NLO level.
These large values are connected with the excitation of the low-lying \(\Delta\)(1232) resonance and can
be understood in terms of the so-called resonance saturation. Indeed, the inclusion of the \(\Delta\) isobar as an
explicit degree of freedom in EFT promotes parts of the 3NFs to NLO \ct*{Epelbaum2008a,Epelbaum2009}.
In systems with strangeness \(S=-1\), resonances such as the \(\Sigma^*\)(1385) could play a similar role as the \(\Delta\) in the \(NNN\) system.
It is therefore likewise compelling to treat their effects in three-baryon forces together with the NLO hyperon-nucleon interaction.

\begin{figure}
\includegraphics[scale=0.6]{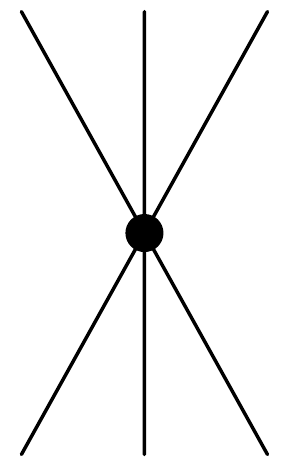}\qquad
\includegraphics[scale=0.6]{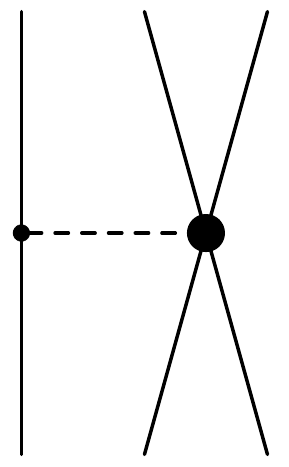}\qquad
\includegraphics[scale=0.6]{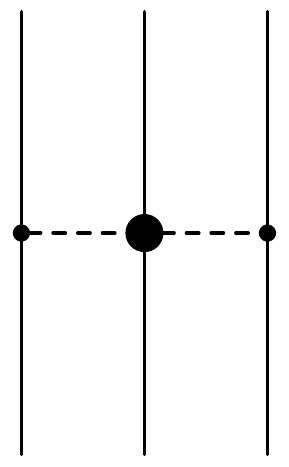}
\caption{
Leading three-baryon interactions: contact term, one-meson exchange and two-meson exchange. Filled circles and solid dots denote vertices with \(\Delta_i=1\) and \(\Delta_i=0\), respectively.
\label{fig:3BF}
}
\end{figure}

In the standard power counting scheme of the baryonic forces in chiral effective field theory (\cf \cts{Epelbaum2009,Machleidt2011}) the chiral dimension \(\nu\) of a given Feynman diagram is determined by
\begin{align*} \label{eq:pwr}
\nu ={}& -4 + 2\mathcal B + 2L + \sum_i v_i \Delta_i\,, \\
&\Delta_i = d_i + \frac12 b_i - 2 \,, \numberthis
\end{align*}
where \(\mathcal B\) is the number of external baryons and \(L\) the number of Goldstone boson loops.
The number of vertices with vertex dimension \(\Delta_i\geq0\) is denoted by \(v_i\).
The symbol \(d_i\) stands for the number of derivatives or pseudoscalar-meson mass insertions at the vertex
and \(b_i\) is the number of internal baryon lines at the considered vertex.
Following \eq{eq:pwr}, one obtains at NNLO with \(\nu=3\) the leading three-baryon diagrams of \fig{fig:3BF} in complete analogy to the leading three-nucleon forces.
Note that a two-meson exchange diagram, like in \fig{fig:3BF}, with a (leading order) Weinberg-Tomozawa vertex in the middle, would formally be a NLO contribution.
However, as in the nucleonic sector, this contribution is kinematically suppressed to higher order.
In SU(3) \cheft{}
nucleons and strange baryons (\(\Lambda\), \(\Sigma\), \(\Xi\)) are treated on equal footing.
Accordingly, reducible diagrams involving those baryons do not constitute genuine three-baryon forces.
These diagrams must not be included into the chiral potential, as they will be generated automatically
when solving the Faddeev or Yakubovsky equations consistently within a coupled-channel approach.
This differs from typical phenomenological calculations with \(\Lambda NN\) three-baryon forces,
where reducible diagrams like the one
with two one-meson exchanges and an intermediate \(\Sigma NN\) state are often used.
In our approach such diagrams do not correspond to a 3BF, but to an iterated two-baryon force.

In this work we construct the potentials for the leading three-baryon forces relevant for few- and many-body calculations, within the framework of SU(3) chiral effective field theory.
The present paper is organized as follows.
In \sect{sec:ct} we show the minimal effective Lagrangian for six-baryon contact terms and its construction principles.
We explain how antisymmetrized potentials can be obtained from the contact Lagrangian.
Furthermore, we investigate the group theoretical classification of the interactions, and provide SU(3) relations for the strangeness 0 and \(-1\) sectors.
In \sect{sec:ome} the minimal chiral Lagrangian for the four-baryon contact vertex involving one pseudoscalar meson is given and applied to the 3BF with one-meson exchange.
\Sect{sec:tme} is devoted to the two-meson exchange potentials.
In \sect{sec:potex} we provide explicit expressions for the potentials of the \(\Lambda NN\) interaction for the contact term and the pion-exchange components.
For comparison the three-body potentials in the nucleonic sector are reproduced.
Conclusions and an outlook are given in \sect{sec:sum}.

\section{Contact interaction} \label{sec:ct}

In the following we consider the three-baryon contact interaction.
We construct the minimal Lagrangian, demonstrate how to derive the antisymmetrized potentials and investigate their group-theoretical classification.

\subsection{Overcomplete contact Lagrangian} \label{subsec:ct1}

The terms of the effective Lagrangian have to fulfill the symmetries of quantum chromodynamics
and are constructed to obey invariance under charge conjugation, parity transformation, Hermitian conjugation and the local chiral symmetry group \(\mathrm{SU}(3)_\mathrm L \times \mathrm{SU}(3)_\mathrm R\).
The baryon fields are collected in the traceless matrix
\begin{equation} \label{eq:baryonmat}
 B=
 \begin{pmatrix}
  \frac{\Sigma^0}{\sqrt 2} + \frac{\Lambda}{\sqrt 6} & \Sigma^+ & p \\
  \Sigma^- & -\frac{\Sigma^0}{\sqrt 2} + \frac{\Lambda}{\sqrt 6} & n \\
  \Xi^- & \Xi^0 & -\frac{2\Lambda}{\sqrt 6}
 \end{pmatrix}\,.
\end{equation}
In order to obtain the most general contact Lagrangian in flavor SU(3), we follow the same procedure as used for the four-baryon contact terms in \ct{Petschauer2013a}.
Generalizing these construction rules straightforwardly to six-baryon contact terms, we end up with a (largely) overcomplete set of terms for the leading covariant Lagrangian:
\begin{equation} \label{eq:Lfull}
\mathcal L = \sum_{f=1}^{11} \sum_{a=1}^5 t^{f,a} \mathcal T^{f,a} \,,
\end{equation}
where the index \(f\) runs over eleven possible flavor structures.
These are given by:
\begin{align*}
 &\mathcal T^{1,a}  ={} \trace{\bBa\bBb\bBg\Ba\Bb\Bg} \\&\ \ + (-1)^{c_a} \trace{\bBg\bBb\bBa\Bg\Bb\Ba}\,, \displaybreak[0]\\
 &\mathcal T^{2,a}  ={} \trace{\bBa\bBb\Ba\bBg\Bb\Bg} \\&\ \ + (-1)^{c_a} \trace{\bBg\bBb\Bg\bBa\Bb\Ba}\,, \displaybreak[0]\\
 &\mathcal T^{3,a}  ={} \trace{\bBa\bBb\Ba\Bb\bBg\Bg} \\&\ \ + (-1)^{c_a} \trace{\bBb\bBa\Bb\Ba\bBg\Bg}\,, \displaybreak[0]\\
 &\mathcal T^{4,a}  ={} \trace{\bBa\Ba\bBb\Bb\bBg\Bg} \\&\ \ + (-1)^{c_a} \trace{\bBg\Bg\bBb\Bb\bBa\Ba}\,, \displaybreak[0]\\
 &\mathcal T^{5,a}  ={} \trace{\bBa\bBb\Ba\Bb}\; \trace{\bBg\Bg} \\&\ \ + (-1)^{c_a} \trace{\bBb\bBa\Bb\Ba}\; \trace{\bBg\Bg}\,, \displaybreak[0]\\
 &\mathcal T^{6,a}  ={} \trace{\bBa\Ba\bBb\Bb}\; \trace{\bBg\Bg} \\&\ \ + (-1)^{c_a} \trace{\bBa\Ba\bBb\Bb}\; \trace{\bBg\Bg}\,, \displaybreak[0]\\
 &\mathcal T^{7,a}  ={} \trace{\bBa\bBb\bBg\Ba}\; \trace{\Bb\Bg} \\&\ \ + (-1)^{c_a} \trace{\bBg\bBb}\; \trace{\bBa\Bg\Bb\Ba}\,, \displaybreak[0]\\
 &\mathcal T^{8,a}  ={} \trace{\bBa\bBb\bBg}\; \trace{\Ba\Bb\Bg} \\&\ \ + (-1)^{c_a} \trace{\bBg\bBb\bBa}\; \trace{\Bg\Bb\Ba}\,, \displaybreak[0]\\
 &\mathcal T^{9,a}  ={} \trace{\bBa\bBb\Ba}\; \trace{\Bb\bBg\Bg} \\&\ \ + (-1)^{c_a} \trace{\bBb\bBg\Bg}\; \trace{\bBa\Bb\Ba}\,, \displaybreak[0]\\
 &\mathcal T^{10,a} ={} \trace{\bBa\Ba}\; \trace{\bBb\Bb}\; \trace{\bBg\Bg} \\&\ \ + (-1)^{c_a} \trace{\bBa\Ba}\; \trace{\bBb\Bb}\; \trace{\bBg\Bg}\,, \displaybreak[0]\\
 &\mathcal T^{11,a} ={} \trace{\bBa\bBb}\; \trace{\bBg\Ba}\; \trace{\Bb\Bg} \\&\ \ + (-1)^{c_a} \trace{\bBg\bBb}\; \trace{\bBa\Bg}\; \trace{\Bb\Ba} \,, \numberthis
\end{align*}
where the indices \(\alpha,\beta,\gamma\) are Dirac indices.
The index \(a=1,\ldots,5\) in \eq{eq:Lfull} labels the three combined Dirac structures \(\Gamma^{1,a}, \Gamma^{2,a}, \Gamma^{3,a}\) that have to be inserted into each flavor structure \(f=1,\dots,11\).
The allowed Dirac structures are given in \tab{tab:BBBgamma}.
Note that we start with a covariant Lagrangian, but in the end are only interested in the minimal non-relativistic Lagrangian.
Therefore, only Dirac structures that lead to independent (non-relativistic) spin operators are considered in \tab{tab:BBBgamma}.
The corresponding spin-dependent potentials \(V_{ijk}^a\) (shown in the last column of \tab{tab:BBBgamma}) are defined by the Dirac structures sandwiched between Dirac spinors in spin spaces \(i\), \(j\) and \(k\).
The overcomplete set of terms in the Lagrangian \eq{eq:Lfull} contains 55 low-energy constants \(t^{f,a}\).
One observes that some combinations of Dirac and flavor structures do not even contribute at the leading order.
Nevertheless, this set is a good starting point to obtain the minimal non-relativistic contact Lagrangian.
\begin{table}
\centering
\begin{tabular}{cccccc}
 \toprule
 \(a\) & \(c_a\) & \(\Gamma^{1,a}\) & \(\Gamma^{2,a}\) & \(\Gamma^{3,a}\) & \( V^a_{ijk} = \) \\\addlinespace[2pt]
 & & & & &\( (\bar u\Gamma^{1,a}u)_i (\bar u\Gamma^{2,a}u)_j (\bar u\Gamma^{3,a}u)_k \) \\
 \cmidrule(lr){1-1} \cmidrule(lr){2-5} \cmidrule(lr){6-6}
 \(1\) & 0 &\(\mathbbm1\) & \(\mathbbm1\) & \(\mathbbm1\) & \(\mathbbm1\) \\
 \(2\) & 0 & \(-\mathbbm1\) & \(\gamma_5\gamma^\mu\) & \(\gamma_5\gamma_\mu\) & \(\vec\sigma_j\cdot\vec\sigma_k\)\\
 \(3\) & 0 & \(\gamma_5\gamma^\mu\) & \(-\mathbbm1\) & \(\gamma_5\gamma_\mu\) & \(\vec\sigma_i\cdot\vec\sigma_k\)\\
 \(4\) & 0 & \(\gamma_5\gamma^\mu\) & \(\gamma_5\gamma_\mu\) & \(-\mathbbm1\) & \(\vec\sigma_i\cdot\vec\sigma_j\)\\
 \(5\) & 1 & \(\gamma_5\gamma_\mu\) & \(-\mathrm i\;\sigma^{\mu\nu}\) & \(\gamma_5\gamma_\nu\) & \(\mathrm i\;\vec\sigma_i\cdot(\vec\sigma_j\times\vec\sigma_k)\)\\
 \bottomrule
\end{tabular}
\caption{Dirac structures \(\Gamma^1, \Gamma^2, \Gamma^3\). Only structures with independent potential contributions are considered. \label{tab:BBBgamma}}
\end{table}

It is advantageous to rewrite the Lagrangian in the particle basis, which gives:
\begin{align*} \label{eq:Lfullpart}
 \mathcal L ={}& \sum_{f=1}^{11} \sum_{a=1}^5 \ \tilde t^{f,a} \sum_{i,j,k,l,m,n} \\ 
 &\times N^{f,a}_{\substack{ikm\\jln}} (\bar B_i\Gamma^{1,a} B_j)(\bar B_k\Gamma^{2,a} B_l)(\bar B_m\Gamma^{3,a} B_n) \,. \numberthis
\end{align*}
where \(B_i\) are the baryon fields in the particle basis and the indices \(i,j,k,l,m,n\) label the six occurring baryon fields, \(B_i\in\{n,p,\Lambda,\Sigma^+,\Sigma^0,\Sigma^-,\Xi^0,\Xi^-\}\).
The SU(3) factors \(N\) can be obtained easily by employing \eq{eq:baryonmat}, multiplying the respective flavor matrices and taking traces.
Note that the constants \(\tilde t^{f,a}\) are equal to \(t^{f,a}\), but with an additional minus sign for \(f=1,3,5,7,8,9,11\), coming from the interchange of anticommuting baryon fields.

\subsection{Derivation of the contact potential}

Let us now consider the process \(B_1 B_2 B_3 \rightarrow B_4 B_5 B_6\), where the \(B_i\) are again baryons in the particle basis.
The aim is to derive a potential operator \(V\) in the (multiple) spin space for this process.
We define the operators in spin-space 1 to act between the two-component Pauli spinors of \(B_1\) and \(B_4\).
Similarly, spin-space 2 belongs to \(B_2\) and \(B_5\), and spin-space 3 to \(B_3\) and \(B_6\).
The potential for a fixed spin configuration is then obtained as
\begin{equation} \label{eq:spinpot}
{\chi_{B_4}^{(1)}}^\dagger {\chi_{B_5}^{(2)}}^\dagger {\chi_{B_6}^{(3)}}^\dagger \, V \, \chi_{B_1}^{(1)}\chi_{B_2}^{(2)}\chi_{B_3}^{(3)} \,,
\end{equation} 
where the superscript of a spinor denotes the spin space and the subscript denotes the baryon to which the spinor belongs.

The potential is given by
\(
V\! =\! -\langle B_4B_5B_6\vert \ \mathcal L \ \vert B_1B_2B_3\rangle
\),
where the appropriate terms of \(\mathcal L\) in \eq{eq:Lfullpart} have to be inserted, and the 36 Wick contractions have to be performed.
First, each of the 55 terms in the Lagrangian (labeled by \(f,a\)) provides six so-called direct terms,
\begin{align*} \label{eq:directerms}
 \phantom{+}\,& \tilde t^{f,a} N^{f,a}_{\substack{456\\123}}(\bar B_4\Gamma^{1,a} B_1)(\bar B_5\Gamma^{2,a} B_2)(\bar B_6\Gamma^{3,a} B_3) \displaybreak[0]\\
           +\,& \tilde t^{f,a} N^{f,a}_{\substack{564\\231}}(\bar B_5\Gamma^{1,a} B_2)(\bar B_6\Gamma^{2,a} B_3)(\bar B_4\Gamma^{3,a} B_1) \displaybreak[0]\\
           +\,& \tilde t^{f,a} N^{f,a}_{\substack{645\\312}}(\bar B_6\Gamma^{1,a} B_3)(\bar B_4\Gamma^{2,a} B_1)(\bar B_5\Gamma^{3,a} B_2) \displaybreak[0]\\
           +\,& \tilde t^{f,a} N^{f,a}_{\substack{465\\132}}(\bar B_4\Gamma^{1,a} B_1)(\bar B_6\Gamma^{2,a} B_3)(\bar B_5\Gamma^{3,a} B_2) \displaybreak[0]\\
           +\,& \tilde t^{f,a} N^{f,a}_{\substack{654\\321}}(\bar B_6\Gamma^{1,a} B_3)(\bar B_5\Gamma^{2,a} B_2)(\bar B_4\Gamma^{3,a} B_1) \displaybreak[0]\\
           +\,& \tilde t^{f,a} N^{f,a}_{\substack{546\\213}}(\bar B_5\Gamma^{1,a} B_2)(\bar B_4\Gamma^{2,a} B_1)(\bar B_6\Gamma^{3,a} B_3) \,, \numberthis
\end{align*}
where the baryon bilinears combine the baryon pairs 1-4, 2-5 and 3-6, in the form as set up in \eq{eq:spinpot}.
Keeping in mind that baryons \(B_1\), \(B_2\), \(B_3\) are in spin-space 1, 2, 3, respectively, one obtains by performing the (six direct) Wick contractions the direct potential%
\footnote{
One observes that \eq{eq:potD} holds independently of whether some of the baryons are identical or not.
}
\begin{align*} \label{eq:potD}
 V^D = -\sum_{f=1}^{11} & \sum_{a=1}^5 {\tilde t}^{f,a} \bigg( 
 N^{f,a}_{\substack{456\\123}}V^a_{123}+ N^{f,a}_{\substack{564\\231}}V^a_{231} + N^{f,a}_{\substack{645\\312}}V^a_{312} \\
 &+ N^{f,a}_{\substack{465\\132}}V^a_{132} + N^{f,a}_{\substack{654\\321}}V^a_{321} + N^{f,a}_{\substack{546\\213}}V^a_{213}
 \bigg) \,. \numberthis
\end{align*}
The spin operators \(V^a_{ijk}\) arise from the Dirac structures \(\Gamma^{1,a} \otimes \Gamma^{2,a} \otimes \Gamma^{3,a}\) and can be found in \tab{tab:BBBgamma}.
The indices \(i,j,k\) of \(V^a_{ijk}\) denote the spin spaces of the three baryon bilinears.

One has not only these six direct Wick contractions, but in total \(36\) Wick contractions that contribute to the potential.
This number corresponds to the \(3!\times3!\) possibilities to arrange the three initial and three final baryons into Dirac bilinears.
For example a term
\begin{equation}
\tilde t^{f,a} N^{f,a}_{\substack{546\\312}}(\bar B_5\Gamma^{1,a} B_3)(\bar B_4\Gamma^{2,a} B_1)(\bar B_6\Gamma^{3,a} B_2)
\end{equation}
gives rise to a potential contribution
\begin{equation}
{\tilde t}^{f,a} N^{f,a}_{\substack{546\\312}} V^a_{312} \,,
\end{equation}
where the sign, reverted in comparison to \eq{eq:potD}, originates from the exchange of baryon fields.
However, this potential is not in accordance with the form of \eq{eq:spinpot}, as baryon pairs 1-4, 2-6, 3-5 are each connected in a separate spin space.
Hence, an exchange of the spin wave functions \(\chi_{B_5}^{(2)}\) and \(\chi_{B_6}^{(3)}\) in the final state  has to be performed and this is achieved by multiplying the potential with \(\Ps_{23}\):
\begin{equation}
{\tilde t}^{f,a} N^{f,a}_{\substack{546\\312}} \Ps_{23} V^a_{312}\,,
\end{equation}
where \(\Ps_{ij}=\frac12(\mathbbm1+\vec\sigma_i\cdot\vec\sigma_j)\) is the well-known spin-exchange operator.

Employing the above considerations to all Wick contractions, the full potential including 36 contributions is derived.
For a shorter notation we express the remaining 30 contributions in terms of the six direct contributions in \eq{eq:potD}, with the declared replacement of the labels.
The full potential is thus given by
\begin{align*} \label{eq:BBBfullpot}
 V &= V^D
 + \Ps_{23}\Ps_{13} \Big(V^D\Big)_{\substack{4\to5\\5\to6\\6\to4}}
 + \Ps_{23}\Ps_{12} \Big(V^D\Big)_{\substack{4\to6\\5\to4\\6\to5}} \\
 &\ - \Ps_{23} \Big(V^D\Big)_{\substack{4\to4\\5\to6\\6\to5}}
 - \Ps_{13} \Big(V^D\Big)_{\substack{4\to6\\5\to5\\6\to4}}
 - \Ps_{12} \Big(V^D\Big)_{\substack{4\to5\\5\to4\\6\to6}} \,. \numberthis
\end{align*}
The procedure described above automatically incorporates the generalized Pauli principle and leads to an antisymmetrized potential.

\subsection{Minimal contact Lagrangian}

Now we are in the position to determine a minimal and complete contact Lagrangian for the leading three-baryon contact interaction.
We have derived the potential according to \eq{eq:BBBfullpot} and decomposed it with respect to the following operators in the three-body spin space
\begin{equation} \label{eq:BBBbasis}
 \mathbbm1\,,\quad
 \vec\sigma_1\cdot\vec\sigma_2\,,\quad
 \vec\sigma_1\cdot\vec\sigma_3\,,\quad
 \vec\sigma_2\cdot\vec\sigma_3\,,\quad
 \mathrm i\;\vec\sigma_1\cdot(\vec\sigma_2\times\vec\sigma_3) \,.
\end{equation}
A minimal set of Lagrangian terms in the non-relativistic limit is obtained by leaving out terms until the rank of the final potential matrix matches the number of terms in the Lagrangian.
Redundant terms have been deleted in such a way, that one obtains a maximal number of Lagrangian terms with a single flavor-trace.
The minimal six-baryon contact Lagrangian in the non-relativistic limit is then given by
\begin{align*} \label{eq:minct}
 \mathcal L =
 -\,&\lc_1 \trace{\bar B_a\bar B_b\bar B_c B_a B_b B_c} \\
 +\,&\lc_2 \trace{\bar B_a\bar B_b B_a\bar B_c B_b B_c} \displaybreak[0]\\
 -\,&\lc_3 \trace{\bar B_a\bar B_b B_a B_b\bar B_c B_c} \displaybreak[0]\\
 +\,&\lc_4 \trace{\bar B_a B_a\bar B_b B_b\bar B_c B_c} \displaybreak[0]\\
 -\,&\lc_5 \trace{\bar B_a\bar B_b B_a B_b}\; \trace{\bar B_c B_c} \displaybreak[0]\\
 -\,&\lc_6 \Big(\trace{\bar B_a\bar B_b\bar B_c B_a(\sigma^i B)_b(\sigma^i B)_c} \\&\qquad + \trace{\bar B_c\bar B_b\bar B_a(\sigma^i B)_c(\sigma^i B)_b B_a}\Big) \displaybreak[0]\\
 +\,&\lc_7 \Big(\trace{\bar B_a\bar B_b B_a\bar B_c(\sigma^i B)_b(\sigma^i B)_c} \\&\qquad + \trace{\bar B_c\bar B_b(\sigma^i B)_c\bar B_a(\sigma^i B)_b B_a}\Big) \displaybreak[0]\\
 -\,&\lc_8 \Big(\trace{\bar B_a\bar B_b B_a(\sigma^i B)_b\bar B_c(\sigma^i B)_c} \\&\qquad + \trace{\bar B_b\bar B_a(\sigma^i B)_b B_a\bar B_c(\sigma^i B)_c}\Big) \displaybreak[0]\\
 +\,&\lc_9 \trace{\bar B_a B_a\bar B_b(\sigma^i B)_b\bar B_c(\sigma^i B)_c} \displaybreak[0]\\
 -\,&\lc_{10} \Big(\trace{\bar B_a\bar B_b B_a(\sigma^i B)_b}\; \trace{\bar B_c(\sigma^i B)_c} \\&\qquad + \trace{\bar B_b\bar B_a(\sigma^i B)_b B_a}\; \trace{\bar B_c(\sigma^i B)_c}\Big) \displaybreak[0]\\
 -\,&\lc_{11} \trace{\bar B_a\bar B_b\bar B_c(\sigma^i B)_a B_b(\sigma^i B)_c} \displaybreak[0]\\
 +\,&\lc_{12} \trace{\bar B_a\bar B_b(\sigma^i B)_a\bar B_c B_b(\sigma^i B)_c} \displaybreak[0]\\
 -\,&\lc_{13} \trace{\bar B_a\bar B_b(\sigma^i B)_a(\sigma^i B)_b\bar B_c B_c} \displaybreak[0]\\
 -\,&\lc_{14} \trace{\bar B_a\bar B_b(\sigma^i B)_a(\sigma^i B)_b}\; \trace{\bar B_c B_c} \displaybreak[0]\\
 -\,&\mathrm i\, \epsilon^{ijk}\lc_{15}\trace{\bar B_a\bar B_b\bar B_c(\sigma^i B)_a(\sigma^j B)_b(\sigma^k B)_c} \displaybreak[0]\\
 +\,&\mathrm i\, \epsilon^{ijk}\lc_{16}\trace{\bar B_a\bar B_b(\sigma^i B)_a\bar B_c(\sigma^j B)_b(\sigma^k B)_c} \displaybreak[0]\\
 -\,&\mathrm i\, \epsilon^{ijk}\lc_{17}\trace{\bar B_a\bar B_b(\sigma^i B)_a(\sigma^j B)_b\bar B_c(\sigma^k B)_c} \\
 +\,&\mathrm i\, \epsilon^{ijk}\lc_{18}\trace{\bar B_a(\sigma^i B)_a\bar B_b(\sigma^j B)_b\bar B_c(\sigma^k B)_c} \,. \numberthis
\end{align*}
The indices \(a,b,c\) are two-component spinor indices and the indices \(i,j,k\) are vector indices.
One ends up with 18 low-energy constants \(\lc_1\dots \lc_{18}\).
The minus signs in front of some terms have been included to compensate minus signs from fermion exchange, arising from reordering baryon bilinears into the form of \eq{eq:Lfullpart}.

Various checks have been performed.
In particular, we verified conservation of strangeness \(S\), isospin \(I\) and isospin projection \(I_3\) and the independence of the resulting potentials from \(I_3\).
The Lagrangian has been constructed to fulfill \(C\) and \(P\) symmetry.
Time reversal symmetry follows via the \(CPT\) theorem, and we explicitly confirmed \(T\) invariance for all potentials.

\subsection{Group-theoretical considerations}

Let us now consider the three-baryon contact terms from a group-theoretical point of view.
In flavor space the three octet baryons form the tensor product \(\ir8 \otimes \ir8 \otimes \ir8\), that decomposes in the following irreducible SU(3) representations
\begin{align*} \label{eq:irrBBB}
\ir8 \otimes \ir8 \otimes \ir8 ={} & \ir{64} \oplus (\ir{35} \oplus \ir{\overline{35}})_2 \oplus \ir{27}_6 \\
& \oplus (\ir{10} \oplus \ir{\overline{10}})_4 \oplus \ir{8}_8 \oplus \ir{1}_2\,, \numberthis
\end{align*}
where the subscripts denote the multiplicity of a representation.
In spin space the tensor product of three doublets decomposes as
\begin{equation}
\ir2 \otimes \ir2 \otimes \ir2 = \ir{2}_2 \oplus \ir{4}\,.
\end{equation}
Transitions are only allowed between irreducible representations of the same type.
In analogy to \ct{Dover1990} for the two-baryon sector, we determine which of the irreducible representations in \eq{eq:irrBBB} can contribute to a particular three-baryon multiplet, characterized by hypercharge \(Y=S+3\) (with strangeness \(S\)) and isospin \(I\).
\Tab{tab:isotuple} gives for the relevant SU(3) representations the \((Y,I)\)-multiplets that they contain.
From this table one can read off which representations are involved in the various three-baryon states, presented in \tab{tab:isoBBB}.
At leading order the potentials are momentum-independent and therefore only \(S\)-waves are present.
Due to the Pauli principle the totally symmetric spin-quartet \(\mathbf4\) must combine with the totally antisymmetric part of \(\ir8 \otimes \ir8 \otimes \ir8\) in flavor space,
\begin{equation}
 \text{Alt}_3(\mathbf8) = \ir{56}_a = \ir{27}_a+\ir{10}_a+\ir{\overline{10}}_a+\ir{8}_a+\ir{1}_a \,.
\end{equation}
Therefore, these totally antisymmetric representations are present only in partial waves with total spin 3/2.
Furthermore, the totally symmetric part of \(\ir8 \otimes \ir8 \otimes \ir8\) decomposes as
\begin{equation}
\text{Sym}_3(\mathbf8) = \ir{120}_s = \ir{64}_s+\ir{27}_s+\ir{10}_s+\ir{\overline{10}}_s+\ir{8}_s+\ir{1}_s \,.
\end{equation}
Since it has no totally antisymmetric counterpart in spin space, it can not contribute.
This is especially true for the highest dimensional \(\ir{64}\) representation, which appears only once in the decomposition \(\ir8 \otimes \ir8 \otimes \ir8\).
In \tab{tab:isoBBB} we have already included these exclusion criteria that follow from the generalized Pauli principle.

\begin{table}
\centering
\begin{tabular}{>{$}c<{$}>{$}c<{$}}
 \toprule
 \quad D\quad \mbox{} & \text{allowed } (Y,I) \\
 \cmidrule(lr){1-1} \cmidrule(lr){2-2} \addlinespace[1pt]
 \mathbf 1 & (0,0) \\\addlinespace[2pt]
 \mathbf 8 & (1,\frac12),(0,0),(0,1),(-1,\frac12)\\\addlinespace[2pt]
 \mathbf{10} & (1,\frac32),(0,1),(-1,\frac12),(-2,0)\\\addlinespace[2pt]
 \mathbf{\overline{10}} & (2,0),(1,\frac12),(0,1),(-1,\frac32)\\\addlinespace[2pt]
 \mathbf{27} & (2,1),(1,\frac12),(1,\frac32),(0,0),(0,1),\\\addlinespace[2pt]
 & (0,2),(-1,\frac12),(-1,\frac32),(-2,1)\\\addlinespace[2pt]
 \mathbf{35} & (2,2),(1,\frac32),(1,\frac52),(0,1),(0,2),(-1,\frac12),\\\addlinespace[2pt]
 & (-1,\frac32),(-2,0),(-2,1),(-3,\frac12) \\\addlinespace[2pt]
 \mathbf{\overline{35}} & (3,\frac12),(2,0),(2,1),(1,\frac12),(1,\frac32),(0,1),\\\addlinespace[2pt]
 & (0,2),(-1,\frac32),(-1,\frac52),(-2,2) \\\addlinespace[2pt]
 \mathbf{64} & (3,\frac32),(2,1),(2,2),(1,\frac12),(1,\frac32),(1,\frac52),\\\addlinespace[2pt]
 & (0,0),(0,1),(0,2),(0,3),(-1,\frac12),\\\addlinespace[2pt]
 & (-1,\frac32),(-1,\frac52),(-2,1),(-2,2),(-3,\frac32) \\\addlinespace[1pt]
 \bottomrule
\end{tabular}
\caption{Hypercharge \(Y\) and isospin \(I\) for irreducible \(SU(3)\) representations of dimension \(D\).
\label{tab:isotuple}
}
\end{table}

\begin{table*}[!ht]
\centering
\begin{tabular}{>{$}c<{$}>{$}c<{$}>{$}c<{$}>{$}c<{$}}
 \toprule
 \text{states} & (Y,I) & {}^2S_{1/2} & {}^4S_{3/2} \\
 \cmidrule(lr){1-2} \cmidrule(lr){3-4}
 NNN & (3,\frac12) & \ir{\overline{35}}\\
 \cmidrule(lr){1-2} \cmidrule(lr){3-4}
 \Lambda NN,\Sigma NN & (2,0) & \ir{\overline{10}},\ir{\overline{35}} & \ir{\overline{10}}_a \\
 \Lambda NN,\Sigma NN & (2,1) & \ir{27},\ir{\overline{35}} & \ir{27}_a \\
 \Sigma NN & (2,2) & \ir{35}\\
 \cmidrule(lr){1-2} \cmidrule(lr){3-4}
 \Lambda\Lambda N,\Sigma\Lambda N,\Sigma\Sigma N,\Xi NN & (1,\frac12) & \ir{8},\ir{\overline{10}},\ir{27},\ir{\overline{35}} & \ir{8}_a,\ir{\overline{10}}_a,\ir{27}_a\\
 \Sigma\Lambda N,\Sigma\Sigma N,\Xi NN & (1,\frac32) & \ir{10},\ir{27},\ir{35},\ir{\overline{35}} & \ir{10}_a,\ir{27}_a\\
 \Sigma\Sigma N & (1,\frac52) & \ir{35} \\
 \cmidrule(lr){1-2} \cmidrule(lr){3-4}
 \Lambda\Lambda\Lambda,\Sigma\Sigma\Lambda,\Sigma\Sigma\Sigma,\Xi\Lambda N,\Xi\Sigma N & (0,0) & \ir{8},\ir{27} & \ir{1}_a,\ir{8}_a,\ir{27}_a\\
 \Sigma\Lambda\Lambda,\Sigma\Sigma\Lambda,\Sigma\Sigma\Sigma,\Xi\Lambda N,\Xi\Sigma N & (0,1) & \ir{8},\ir{10},\ir{\overline{10}},\ir{27},\ir{35},\ir{\overline{35}} & \ir{8}_a,\ir{10}_a,\ir{\overline{10}}_a,\ir{27}_a\\
 \Sigma\Sigma\Lambda,\Sigma\Sigma\Sigma,\Xi\Sigma N & (0,2) & \ir{27},\ir{35},\ir{\overline{35}} & \ir{27}_a \\
 \cmidrule(lr){1-2} \cmidrule(lr){3-4}
 \Xi\Lambda\Lambda,\Xi\Sigma\Lambda,\Xi\Sigma\Sigma,\Xi\Xi N & (-1,\frac12) & \ir{8},\ir{10},\ir{27},\ir{35} & \ir{8}_a,\ir{10}_a,\ir{27}_a\\
 \Xi\Sigma\Lambda,\Xi\Sigma\Sigma,\Xi\Xi N & (-1,\frac32) & \ir{\overline{10}},\ir{27},\ir{35},\ir{\overline{35}} & \ir{\overline{10}}_a,\ir{27}_a \\
 \Xi\Sigma\Sigma & (-1,\frac52) & \ir{\overline{35}} \\
 \cmidrule(lr){1-2} \cmidrule(lr){3-4}
 \Xi\Xi\Lambda,\Xi\Xi\Sigma & (-2,0) & \ir{10},\ir{35} & \ir{10}_a\\
 \Xi\Xi\Lambda,\Xi\Xi\Sigma & (-2,1) & \ir{27},\ir{35} & \ir{27}_a \\
 \Xi\Xi\Sigma & (-2,2) & \ir{\overline{35}}\\
 \cmidrule(lr){1-2} \cmidrule(lr){3-4}
 \Xi\Xi\Xi & (-3,\frac12) & \ir{35}\\
 \bottomrule
\end{tabular}
\caption{Irreducible representations for three-baryon states with hypercharge \(Y\) and isospin \(I\) in partial waves.
\label{tab:isoBBB}
}
\end{table*}

In the next step, we can derive the potentials for transitions between the three-baryon states, and redefine the 18 constants such that they belong to transitions between irreducible representations.
It is a highly non-trivial check of our results that this redefinition meets the restrictions of \tab{tab:isoBBB}.
For example, in the \(NNN\) interaction and the \(\Xi\Xi\Sigma\ (-2,2)\) interaction the same constant associated with the \(\ir{\overline{35}}\) representation has to be present.

In order to obtain a representation of the potentials in the isospin basis, we use the relation%
\footnote{
In order to obtain \tab{tab:PWDBBB} we strictly employ \eq{eq:matelemiso33}, \ie no further combinatorial factors, such as \(1/\sqrt2\) for a \(\Lambda NN\) state are included.
They can be included by just multiplying the corresponding row in \tab{tab:PWDBBB} with that factor.
}
\begin{align*} \label{eq:matelemiso33}
 &\langle(i_4i_5)i_\mathrm{out}(i_\mathrm{out}i_6)I_\mathrm{out}M_\mathrm{out}|\hat{\mathcal O}|(i_1i_2)i_\mathrm{in}(i_\mathrm{in}i_3)I_\mathrm{in}M_\mathrm{in}\rangle \\
 & =
 \sum_{\substack{m_1,m_2,m_3,m_\mathrm{in},\\m_4,m_5,m_6,m_\mathrm{out}}}
 \delta_{m_\mathrm{out},m_4+m_5}
 \delta_{M_\mathrm{out},m_\mathrm{out}+m_6}\\
 &\qquad \times
 \delta_{m_\mathrm{in},m_1+m_2}
 \delta_{M_\mathrm{in},m_\mathrm{in}+m_3}\\
 &\qquad \times
 C^{i_4i_5i_\mathrm{out}}_{m_4m_5m_\mathrm{out}} C^{i_\mathrm{out}i_6I_\mathrm{out}}_{m_\mathrm{out}m_6M_\mathrm{out}}
 C^{i_1i_2i_\mathrm{in}}_{m_1m_2m_\mathrm{in}}
 C^{i_\mathrm{in}i_3I_\mathrm{in}}_{m_\mathrm{in}m_3M_\mathrm{in}} \\
 &\qquad \times
 \langle i_4m_4;i_5m_5;i_6m_6|\hat{\mathcal O}|i_1m_1;i_2m_2;i_3m_3\rangle\,, \numberthis
\end{align*}
where \(i\) stands for the isospin and \(C\) are the Clebsch-Gordan coefficients.
In order to be consistent with the Condon-Shortley convention for the Clebsch-Gordan coefficients, we use the baryon matrix as defined in \eq{eq:baryonmat} and make the following sign changes in the identification of the particle states \(\vert i,m\rangle \):
\(\Sigma^+=-\vert 1,+1\rangle\,,\ \Xi^-=-\vert 1/2,-1/2\rangle\).
In \eq{eq:matelemiso33} we have chosen to couple the isospin of the first two particle in the initial state \(i_1\), \(i_2\) to \(i_\mathrm{in}\) and then to couple \(i_\mathrm{in}\) with the isospin \(i_3\) of the third particle to total isospin \(I_\mathrm{in}\).
The same procedure is applied to the final state.
Other coupling schemes can be obtained by recoupling with the help of Racah \(W\)-coefficients or equivalently with Wigners 6\(j\)-symbols.

It is advantageous to present the three-body potentials not only in terms of the spin operators in \eq{eq:BBBbasis}, but to project them also onto partial wave contributions.
For a general operator
\begin{align*}
\hat{\mathcal O} = {} &
a_1\, \mathbbm1+
a_2\, \vec\sigma_1\cdot\vec\sigma_2+
a_3\, \vec\sigma_1\cdot\vec\sigma_3 \\
&+a_4\, \vec\sigma_2\cdot\vec\sigma_3+
a_5\, \mathrm i\, \vec\sigma_1\times\vec\sigma_2\cdot\vec\sigma_3 \,, \numberthis
\end{align*}
with coefficients \(a_i\), the partial wave decomposition leads to the following non-vanishing transitions (between \(S\)-waves):
\begin{align*} 
\langle 0\ {}^{2}S_{1/2} \vert \hat{\mathcal O} \vert 0\ {}^{2}S_{1/2} \rangle &= a_1-3a_2 \,, \\
\langle 1\ {}^{2}S_{1/2} \vert \hat{\mathcal O} \vert 0\ {}^{2}S_{1/2} \rangle &= \sqrt3(- a_3 +  a_4 - 2 a_5) \,, \\
\langle 0\ {}^{2}S_{1/2} \vert \hat{\mathcal O} \vert 1\ {}^{2}S_{1/2} \rangle &= \sqrt3(- a_3 +  a_4 + 2 a_5) \,, \\
\langle 1\ {}^{2}S_{1/2} \vert \hat{\mathcal O} \vert 1\ {}^{2}S_{1/2} \rangle &= a_1 + a_2 - 2a_3 - 2a_4 \,, \\
\langle 1\ {}^{4}S_{3/2} \vert \hat{\mathcal O} \vert 1\ {}^{4}S_{3/2} \rangle &= a_1+a_2+a_3+a_4 \,, \numberthis
\end{align*}
where a state \(\vert s\ {}^{2S+1}L_J \rangle\) is characterized by the total spin \(S=\frac12,\frac32\), the angular momentum \(L=0\) and the total angular momentum \(J=\frac12,\frac32\).
Here, we have chosen to couple the spins of the first two baryons to \(s=0,1\), and to couple this with the spin \(\frac12\) of the third baryon to \(S\) (in complete analogy to the  isospin coupling in \eq{eq:matelemiso33}).
After this partial wave decomposition it is trivial to identify the combinations of constants belonging to the totally antisymmetric flavor representations, since these act only in the \(1\ {}^{4}S_{3/2}\) states due to the generalized Pauli principle.

Finally, we give the SU(3) relations for the strangeness \(0\) and \(-1\) sectors in \tab{tab:PWDBBB}.
The constants associated with the irreducible SU(3) representations are related to the low-energy constants of the minimal Lagrangian by:
\begin{widetext}
\vspace{-1.2\baselineskip}
\begin{align*} \label{eq:groupconst}
c_{\overline{35}} &= 6 (-\lc_4 +\lc_9 ) \,, \\ 
c_{35} &= 3 (\lc_4 -\lc_9 + 6 \lc_{18} ) \,, \displaybreak[0]\\ 
c_{\overline{10}} &= \frac{3}{4} (2 \lc_2 +\lc_3 -\lc_4 +\lc_5 -6 \lc_8 +\lc_9  - 6 \lc_{10} -6 \lc_{12} +3 \lc_{13} +3 \lc_{14} +6 \lc_{17} -6 \lc_{18} ) \,, \displaybreak[0]\\ 
c_{27^1} &= -\frac{37 }{294} \lc_2+\frac{769 }{588} \lc_3-\frac{473 }{392} \lc_4+\frac{769 }{588} \lc_5-\frac{74 }{49} \lc_7-\frac{429 }{98} \lc_8+\frac{473 }{392} \lc_9  \displaybreak[0]\\
 &\phantom{ = \ }  - \frac{429 }{98} \lc_{10}+\frac{185 }{98} \lc_{12}+\frac{89 }{196} \lc_{13}+\frac{89 }{196} \lc_{14}+\frac{244 }{49} \lc_{16}-\frac{207 }{98} \lc_{17}+\frac{57}{14} \lc_{18} \,, \displaybreak[0]\\ 
c_{27^2} &= \frac{1}{24} (-4 \lc_2 -22 \lc_3 +57 \lc_4 -22 \lc_5 -48 \lc_7 -12 \lc_8 -57 \lc_9 - 12 \lc_{10} \\
 &\phantom{ = \ }   +60 \lc_{12} +78 \lc_{13} +78 \lc_{14} -96 \lc_{16} +60 \lc_{17} -252 \lc_{18} ) \,, \displaybreak[0]\\ 
c_{27^3} &= \frac{1}{8} ( 20 \lc_2 -2 \lc_3 -21 \lc_4 -2 \lc_5 -16 \lc_7 +28 \lc_8 +21 \lc_9  +28 \lc_{10} -44 \lc_{12} \\
 &\phantom{ = \ } -22 \lc_{13} -22 \lc_{14} +32 \lc_{16} -76 \lc_{17} +12 \lc_{18}) \,, \displaybreak[0]\\ 
c_{\overline{10}a} &= 6 (  -\lc_2 +\lc_3 -\lc_4 +\lc_5 -2 \lc_7 +2 \lc_8 -\lc_9 + 2 \lc_{10} -\lc_{12} +\lc_{13} +\lc_{14} ) \,, \displaybreak[0]\\ 
c_{27a} &= \frac{2}{3} (\lc_2 +\lc_3 +3 \lc_4 +\lc_5 +2 \lc_7 +2 \lc_8 +3 \lc_9 + 2 \lc_{10} +\lc_{12} +\lc_{13} +\lc_{14} )
\,.\numberthis
\end{align*}
\end{widetext}
The SU(3) relations have not been obtained by group theory considerations directly, but by rewriting our results such that they fulfill the group-theoretical constraints of \tab{tab:isoBBB}.
The three constants \(\lc_{27^1}\), \(\lc_{27^2}\), \(\lc_{27^3}\) are associated to the irreducible representations of dimension \(\ir{27}\).
We have chosen a particular definition for them in \eq{eq:groupconst}.
Note that other linear combinations of \(\lc_{27^1}\), \(\lc_{27^2}\), \(\lc_{27^3}\) would work equally well.
The SU(3) relations in \tab{tab:PWDBBB} have been derived from the most general SU(3) symmetric Lagrangian.
Therefore, any three-baryon potential that fulfills flavor SU(3) symmetry has to fulfill these relations.
These relations provide also a valuable check for the SU(3) decomposition of the \(S\)-wave contributions from three-baryon interactions generated by one- or two-meson exchange (with all meson masses set equal).

\afterpage{\afterpage{
\newcolumntype{f}{>{$}c<{$}}
\newcolumntype{F}{>{\centering\arraybackslash$}p{3.5cm}<{$}}
\begin{turnpage}
\begin{table}
\centering
\begin{tabular}{ffffFFFFF}
\toprule
\text{transition} & I & i_{in} & i_{out} & V_{0 ^2S_{1/2}\rightarrow 0 {}^2S_{1/2}} & V_{0 {}^2S_{1/2}\rightarrow 1 {}^2S_{1/2}} & V_{1 {}^2S_{1/2}\rightarrow 0 {}^2S_{1/2}} & V_{1 {}^2S_{1/2}\rightarrow 1 {}^2S_{1/2}} & V_{1 {}^4S_{3/2}\rightarrow 1 {}^4S_{3/2}}\\ 
\cmidrule(lr){1-4}\cmidrule(lr){5-9}
N N N \to N N N & \frac{1}{2} & 0 & 0 & 0 & 0 & 0 & 3 c_{\overline{35}} & 0 \\ 
N N N \to N N N & \frac{1}{2} & 1 & 0 & 0 & -3 c_{\overline{35}} & 0 & 0 & 0 \\ 
N N N \to N N N & \frac{1}{2} & 0 & 1 & 0 & 0 & -3 c_{\overline{35}} & 0 & 0 \\ 
N N N \to N N N & \frac{1}{2} & 1 & 1 & 3 c_{\overline{35}} & 0 & 0 & 0 & 0 \\ 

\cmidrule(lr){1-4}\cmidrule(lr){5-9}
\Lambda  N N \to \Lambda  N N & 0 & \frac{1}{2} & \frac{1}{2} & c_{\overline{10}}+c_{\overline{35}} & \frac{c_{\overline{10}}}{\sqrt{3}}+\frac{c_{\overline{35}}}{\sqrt{3}} & \frac{c_{\overline{10}}}{\sqrt{3}}+\frac{c_{\overline{35}}}{\sqrt{3}} & \frac{c_{\overline{10}}}{3}+\frac{c_{\overline{35}}}{3} & c_{\overline{10}a} \\ 
\Lambda  N N \to \Lambda  N N & 1 & \frac{1}{2} & \frac{1}{2} & c_{27^1}+\frac{12 c_{27^2}}{49}+\frac{3 c_{\overline{35}}}{16} & -\sqrt{3} c_{27^1}-\frac{12 \sqrt{3} c_{27^2}}{49}-\frac{3 \sqrt{3} c_{\overline{35}}}{16} & -\sqrt{3} c_{27^1}-\frac{12 \sqrt{3} c_{27^2}}{49}-\frac{3 \sqrt{3} c_{\overline{35}}}{16} & 3 c_{27^1}+\frac{36 c_{27^2}}{49}+\frac{9 c_{\overline{35}}}{16} & 0 \\ 
\Sigma  N N \to \Sigma  N N & 0 & \frac{1}{2} & \frac{1}{2} & c_{\overline{10}} & -\sqrt{3} c_{\overline{10}} & -\sqrt{3} c_{\overline{10}} & 3 c_{\overline{10}} & 0 \\ 
\Sigma  N N \to \Sigma  N N & 1 & \frac{1}{2} & \frac{1}{2} & c_{27^2}+\frac{c_{\overline{35}}}{48} & \frac{c_{27^3}}{\sqrt{3}}+\frac{\sqrt{3} c_{\overline{35}}}{16} & \frac{c_{27^3}}{\sqrt{3}}+\frac{\sqrt{3} c_{\overline{35}}}{16} & \frac{4 c_{27^1}}{3}-\frac{c_{27^2}}{147}-\frac{2 c_{27^3}}{3}+\frac{9 c_{\overline{35}}}{16} & c_{27a} \\ 
\Sigma  N N \to \Sigma  N N & 1 & \frac{3}{2} & \frac{1}{2} & -\frac{5 c_{27^2}}{4 \sqrt{2}}-\frac{3 c_{27^3}}{4 \sqrt{2}}-\frac{c_{\overline{35}}}{6 \sqrt{2}} & -\sqrt{\frac{3}{2}} c_{27^1}+\frac{1}{196} \sqrt{\frac{3}{2}} c_{27^2}+\frac{c_{27^3}}{4 \sqrt{6}}-\frac{1}{2} \sqrt{\frac{3}{2}} c_{\overline{35}} & \frac{c_{27^3}}{4 \sqrt{6}}-\frac{3}{4} \sqrt{\frac{3}{2}} c_{27^2} & \frac{c_{27^1}}{3 \sqrt{2}}-\frac{c_{27^2}}{588 \sqrt{2}}-\frac{11 c_{27^3}}{12 \sqrt{2}} & -\sqrt{2} c_{27a} \\ 
\Sigma  N N \to \Sigma  N N & 1 & \frac{1}{2} & \frac{3}{2} & -\frac{5 c_{27^2}}{4 \sqrt{2}}-\frac{3 c_{27^3}}{4 \sqrt{2}}-\frac{c_{\overline{35}}}{6 \sqrt{2}} & \frac{c_{27^3}}{4 \sqrt{6}}-\frac{3}{4} \sqrt{\frac{3}{2}} c_{27^2} & -\sqrt{\frac{3}{2}} c_{27^1}+\frac{1}{196} \sqrt{\frac{3}{2}} c_{27^2}+\frac{c_{27^3}}{4 \sqrt{6}}-\frac{1}{2} \sqrt{\frac{3}{2}} c_{\overline{35}} & \frac{c_{27^1}}{3 \sqrt{2}}-\frac{c_{27^2}}{588 \sqrt{2}}-\frac{11 c_{27^3}}{12 \sqrt{2}} & -\sqrt{2} c_{27a} \\ 
\Sigma  N N \to \Sigma  N N & 1 & \frac{3}{2} & \frac{3}{2} & \frac{9 c_{27^1}}{8}+\frac{38 c_{27^2}}{49}+\frac{3 c_{27^3}}{8}+\frac{2 c_{\overline{35}}}{3} & -\frac{\sqrt{3} c_{27^1}}{8}+\frac{23 \sqrt{3} c_{27^2}}{49}+\frac{7 c_{27^3}}{8 \sqrt{3}} & -\frac{\sqrt{3} c_{27^1}}{8}+\frac{23 \sqrt{3} c_{27^2}}{49}+\frac{7 c_{27^3}}{8 \sqrt{3}} & \frac{c_{27^1}}{24}+\frac{124 c_{27^2}}{147}-\frac{5 c_{27^3}}{24} & 2 c_{27a} \\ 
\Sigma  N N \to \Sigma  N N & 2 & \frac{3}{2} & \frac{3}{2} & \frac{c_{35}}{2} & -\frac{\sqrt{3} c_{35}}{2} & -\frac{\sqrt{3} c_{35}}{2} & \frac{3 c_{35}}{2} & 0 \\ 
\Lambda  N N \to \Sigma  N N & 0 & \frac{1}{2} & \frac{1}{2} & c_{\overline{10}}-\frac{c_{\overline{35}}}{2} & \frac{\sqrt{3} c_{\overline{35}}}{2}-\sqrt{3} c_{\overline{10}} & \frac{c_{\overline{10}}}{\sqrt{3}}-\frac{c_{\overline{35}}}{2 \sqrt{3}} & \frac{c_{\overline{35}}}{2}-c_{\overline{10}} & 0 \\ 
\Lambda  N N \to \Sigma  N N & 1 & \frac{1}{2} & \frac{1}{2} & \frac{c_{27^2}}{2}+\frac{c_{27^3}}{2}-\frac{c_{\overline{35}}}{16} & \frac{2 c_{27^1}}{\sqrt{3}}-\frac{c_{27^2}}{98 \sqrt{3}}-\frac{c_{27^3}}{2 \sqrt{3}}-\frac{3 \sqrt{3} c_{\overline{35}}}{16} & -\frac{\sqrt{3} c_{27^2}}{2}-\frac{\sqrt{3} c_{27^3}}{2}+\frac{\sqrt{3} c_{\overline{35}}}{16} & -2 c_{27^1}+\frac{c_{27^2}}{98}+\frac{c_{27^3}}{2}+\frac{9 c_{\overline{35}}}{16} & 0 \\ 
\Lambda  N N \to \Sigma  N N & 1 & \frac{1}{2} & \frac{3}{2} & -\frac{3 c_{27^1}}{2 \sqrt{2}}-\frac{121 c_{27^2}}{196 \sqrt{2}}-\frac{c_{27^3}}{4 \sqrt{2}}+\frac{c_{\overline{35}}}{2 \sqrt{2}} & \frac{c_{27^1}}{2 \sqrt{6}}-\frac{221 c_{27^2}}{196 \sqrt{6}}-\frac{5 c_{27^3}}{4 \sqrt{6}} & \frac{3}{2} \sqrt{\frac{3}{2}} c_{27^1}+\frac{121}{196} \sqrt{\frac{3}{2}} c_{27^2}+\frac{1}{4} \sqrt{\frac{3}{2}} c_{27^3}-\frac{1}{2} \sqrt{\frac{3}{2}} c_{\overline{35}} & -\frac{c_{27^1}}{2 \sqrt{2}}+\frac{221 c_{27^2}}{196 \sqrt{2}}+\frac{5 c_{27^3}}{4 \sqrt{2}} & 0 \\ 

\bottomrule
\end{tabular}
\caption{SU(3) relations of three-baryon contact terms with strangeness \(0\) and \(-1\) in non-vanishing partial waves.\label{tab:PWDBBB}}
\end{table}
\end{turnpage}
\clearpage
}}

\section{One-meson exchange} \label{sec:ome}

For the one-meson exchange diagram in \fig{fig:3BF} we employ the standard chiral Lagrangian for meson-baryon couplings \ct*{Bernard1995}%
\begin{equation} \label{eq:LBBM}
\mathcal{L} = \frac D 2 \langle \bar B \gamma^\mu \gamma_5 \lbrace u_\mu,B\rbrace\rangle + \frac F 2 \langle\bar B \gamma^\mu \gamma_5 \left[u_\mu,B\right]\rangle \,,
\end{equation}
with the axial vector coupling constants \(D\approx0.8\) and \(F\approx0.5\) and \(u_\mu = -\frac1{f_0}\partial_\mu\phi+\mathcal O(\phi^3)\), where the pseudoscalar-meson fields are collected in the traceless Hermitian matrix
\begin{equation} \label{eq:mesonmat}
\phi=
\begin{pmatrix}
\pi^0 + \frac{\eta}{\sqrt 3} & \sqrt 2 \pi^+ & \sqrt 2 K^+ \\
\sqrt 2 \pi^- & -\pi^0 + \frac{\eta}{\sqrt 3} & \sqrt 2 K^0 \\
\sqrt 2 K^- & \sqrt 2 \bar K^0 & -\frac{2\eta}{\sqrt 3}
\end{pmatrix} \,.
\end{equation}
Here, \(f_0\) is the pion decay constant (in the chiral limit). 
As done in \eq{eq:Lfullpart} it is advantageous to express this Lagrangian in the particle basis:
\begin{equation}
 \mathcal{L}=- \sum_{i,j,k} \frac{1}{2f_0} N_{B_iB_j\phi_k} (\bar B_i \gamma^\mu \gamma_5 B_j) (\partial_\mu \phi_k)\,,
\end{equation}
with the baryon fields as defined before \(B_i\in\{n,\linebreak[0]p,\linebreak[0]\Lambda,\linebreak[0]\Sigma^+,\linebreak[0]\Sigma^0,\linebreak[0]\Sigma^-,\linebreak[0]\Xi^0,\linebreak[0]\Xi^-\}\) and the pseudoscalar-meson fields
\(\phi_i \in \left\{\pi^0,\linebreak[0]\pi^+,\linebreak[0]\pi^-,\linebreak[0]K^+,\linebreak[0]K^-,\linebreak[0]K^0,\linebreak[0]\bar K^0,\linebreak[0]\eta\right\} \).

The second vertex, necessary for the one-meson-exchange three-body interaction, involves four baryon fields and one pseudoscalar-meson field.
An overcomplete set of terms for the corresponding relativistic Lagrangian can be found in our earlier work, \ct{Petschauer2013a}.
In order to obtain a complete minimal set of terms in the non-relativistic limit, we consider the matrix elements of the process \(B_1B_2\to B_3B_4\phi_1\) and proceed as in \sect{sec:ct}.
The transition matrix element is expressed in terms of the spin operators
\begin{equation}
\vec\sigma_1\cdot\vec q\,,\quad
\vec\sigma_2\cdot\vec q\,,\quad
\mathrm i\,(\vec\sigma_1\times\vec\sigma_2)\cdot\vec q \,,
\end{equation}
where \(\vec q\) denotes the momentum of the emitted meson.
The minimal Lagrangian is obtained by eliminating redundant terms until the rank of the matrix formed by all transitions matches the number of terms in the Lagrangian.
As before, redundant terms are deleted in such a way, that one obtains a maximal number of terms with a single flavor trace.
The minimal non-relativistic chiral Lagrangian for the four-baryon vertex including one meson is given by
\begin{align*} \label{eq:LBBMBBmin}
 \mathcal L ={}
 &\ld_1/f_0 \trace{\bar B_a \extfield B_a\bar B_b(\sigma^i B)_b}\\
 &+\ld_2/f_0 \Big( \trace{\bar B_a  B_a \extfield\bar B_b(\sigma^i B)_b} \\&\qquad\qquad + \trace{\bar B_a  B_a\bar B_b(\sigma^i B)_b \extfield}\Big)\displaybreak[0]\\
 &+\ld_3/f_0 \trace{\bar B_b \extfield(\sigma^i B)_b\bar B_a B_a}\displaybreak[0]\\
 &-\ld_4/f_0 \Big( \trace{\bar B_a\extfield\bar B_b B_a(\sigma^i B)_b} \\&\qquad\qquad + \trace{\bar B_b \bar B_a (\sigma^i B)_b\extfield B_a}\Big)\displaybreak[0]\\
 &-\ld_5/f_0 \Big( \trace{\bar B_a\bar B_b\extfield B_a(\sigma^i B)_b} \\&\qquad\qquad + \trace{\bar B_b \bar B_a \extfield (\sigma^i B)_b B_a}\Big)\displaybreak[0]\\
 &-\ld_6/f_0 \Big( \trace{\bar B_b\extfield\bar B_a(\sigma^i B)_b B_a} \\&\qquad\qquad + \trace{\bar B_a \bar B_b B_a\extfield (\sigma^i B)_b}\Big)\displaybreak[0]\\
 &-\ld_7/f_0 \Big( \trace{\bar B_a\bar B_b B_a(\sigma^i B)_b\extfield} \\&\qquad\qquad + \trace{\bar B_b \bar B_a (\sigma^i B)_b B_a\extfield}\Big)\displaybreak[0]\\
 &+\ld_8/f_0 \trace{\bar B_a\extfield B_a}\trace{\bar B_b(\sigma^i B)_b}\displaybreak[0]\\
 &+\ld_9/f_0 \trace{\bar B_a B_a\extfield}\trace{\bar B_b(\sigma^i B)_b}\displaybreak[0]\\
 &+\ld_{10}/f_0 \trace{\bar B_b\extfield(\sigma^i B)_b}\trace{\bar B_a B_a}\displaybreak[0]\\
 &+\mathrm i\,\epsilon^{ijk}\ld_{11}/f_0 \trace{\bar B_a (\sigma^i B)_a (\nabla^k \phi)\bar B_b(\sigma^j B)_b}\displaybreak[0]\\
 &-\mathrm i\,\epsilon^{ijk}\ld_{12}/f_0 \Big( \trace{\bar B_a(\nabla^k \phi)\bar B_b(\sigma^i B)_a(\sigma^j B)_b} \\&\qquad\qquad\qquad\ - \trace{\bar B_b \bar B_a (\sigma^j B)_b(\nabla^k \phi) (\sigma^i B)_a}\Big)\displaybreak[0]\\
 &-\mathrm i\,\epsilon^{ijk}\ld_{13}/f_0 \trace{\bar B_a\bar B_b(\nabla^k \phi)(\sigma^i B)_a(\sigma^j B)_b}\\
 &-\mathrm i\,\epsilon^{ijk}\ld_{14}/f_0 \trace{\bar B_a\bar B_b(\sigma^i B)_a(\sigma^j B)_b(\nabla^k \phi)}\,. \numberthis
\end{align*}
Here, the indices \(a\) and \(b\) are two-component spinor indices and the indices \(i\), \(j\) and \(k\) are vector indices.
There are in total 14 low-energy constants \(\ld_1\dots \ld_{14}\) for all five strangeness sectors \(S=-4\ldots0\).
As before, the minus signs in front of some terms have been included, in order to compensate minus signs from fermion exchange, arising from reordering baryon bilinears (see \eq{eq:LBBMBBpart} below).
Let us note, that the conservation of strangeness \(S\), isospin \(I\) and isospin projection \(I_3\), independence of \(I_3\), and time reversal symmetry have been checked for the \(BB\to BB\phi\) transition matrix elements resulting from \eq{eq:LBBMBBmin}.
Moreover, several tests employing group theoretical methods have been performed.

As done in \sect{subsec:ct1}, we write the Lagrangian in the particle basis
\begin{align*} \label{eq:LBBMBBpart}
\mathcal L ={}& \quad \sum_{f=1}^{10} \frac{\ld_f}{f_0} \sum_{i,j,k,l,m=1}^{8} N^{f}_{\substack{ik\\jl}\,\phi_m} \\
& \qquad \times (\bar B_i B_j)(\bar B_k\vec\sigma B_l)\cdot \vec\nabla\phi_m \displaybreak[0]\\
& +\sum_{f=11}^{14} \frac{\ld_f}{f_0} \sum_{i,j,k,l,m=1}^{8} N^{f}_{\substack{ik\\jl}\,\phi_m} \\
& \qquad \times \mathrm i\,[(\bar B_i \vec\sigma B_j)\times(\bar B_k\vec\sigma B_l) ]\cdot \vec\nabla\phi_m \,, \numberthis
\end{align*}
where in each term the first bilinear comes from the summation over spin index \(a\) and the second bilinear from the summation over spin index \(b\) in \eq{eq:LBBMBBmin}.
The indices \(i,j,k,l\) label octet baryons.

\begin{figure}
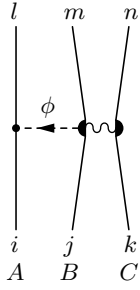

\vspace{.3\baselineskip}
\centering
\begin{overpic}[scale=.6]{FBBB1ME-dot2}
\put(2,101){$l$}\put(27,101){$m$}\put(55,101){$n$}
\put(2,-8){$i$}\put(27,-8){$j$}\put(55,-8){$k$}
\put(0,-21){$A$}\put(25,-21){$B$}\put(53,-21){$C$}
\put(16,57){$\phi$}
\end{overpic}
\vspace{.8\baselineskip}
\caption{
Generic one-meson exchange diagram. The wiggly line symbolized the four-baryon contact vertex, to illustrate the baryon bilinears.
\label{fig:ome-gen}
}
\end{figure}

Let us now consider the generic one-meson exchange diagram in \fig{fig:ome-gen}.
It involves the baryons \(i,j,k\) in the initial state, the baryons \(l,m,n\) in the final state and an exchanged meson \(\phi\).
The four-baryon contact vertex is separated into two parts, in order to indicate which baryons are in the same bilinear.
The indices \(A,B,C\) label the spin spaces related to the baryon bilinears.

Using standard Feynman rules for the vertices and the meson propagator one obtains the following three-body potential
\begin{equation}
V = \frac{1}{2f_0^2} \frac{\vec\sigma_A\cdot\vec q_{li}}{\vec q_{li}^{\,2}+m_{\phi}^2} \Big(
N_1 \vec\sigma_C\cdot\vec q_{li}
+N_2 \mathrm i\,(\vec\sigma_B\times\vec\sigma_C)\cdot\vec q_{li}
\Big)\,,
\end{equation}
with the momentum transfer \(\vec q_{li} = \vec p_l-\vec p_i\) carried by the exchanged meson and the constants
\begin{align*}
N_1 &= N_{B_lB_i\phi} \sum_{f=1}^{10} \ld_f N^{f}_{\substack{mn\\jk}\,\bar\phi} \ , \\
N_2 &= N_{B_lB_i\phi} \sum_{f=11}^{14} \ld_f N^{f}_{\substack{mn\\jk}\,\bar\phi} \ , \numberthis
\end{align*}
where \(\bar\phi\) denotes the charge-conjugated meson of meson \(\phi\), in particle basis (\eg \(\pi^+\leftrightarrow\pi^-\)).

The full one-meson exchange three-body potential for the process \(B_1B_2B_3\to B_4B_5B_6\) is obtained easily by summing up for a fixed meson the 36 permutations of initial and final baryons, shown diagrammatically in \fig{fig:omefeyn}, and summing over all mesons \(\phi \in \left\{\pi^0,\pi^+,\pi^-,K^+,K^-,K^0,\bar K^0,\eta\right\}\).
Of course, many of these contributions will vanish for a particular process.
The Feynman diagrams fall into 9 classes, where in each class the same momentum transfer \(\vec q_{li}\) is present.
In \fig{fig:omefeyn} each row corresponds to such a class and the corresponding momentum transfer is written on the left of the row.
Furthermore, additional minus signs from interchanging fermions have to be included and some diagrams need to be multiplied from the left by spin exchange operators (as indicated in \fig{fig:omefeyn}) in order to be in accordance with the form set up in \eq{eq:spinpot}.
As before, the baryons \(B_1\), \(B_2\) and \(B_3\) belong to the spin-spaces 1, 2 and 3, respectively.

\afterpage{
\begin{figure*}
\begin{alignat*}{2}
&\vec q_{41}\colon&&
\vcenter{\hbox{\BBBOMEpic{B_1}{B_2}{B_3}{B_4}{B_5}{B_6}}} +
\vcenter{\hbox{\BBBOMEpic{B_1}{B_3}{B_2}{B_4}{B_6}{B_5}}}
-\Ps_{23}\Bigg(
\vcenter{\hbox{\BBBOMEpic{B_1}{B_2}{B_3}{B_4}{B_6}{B_5}}} +
\vcenter{\hbox{\BBBOMEpic{B_1}{B_3}{B_2}{B_4}{B_5}{B_6}}}
\Bigg) \\
&\vec q_{52}\colon&&
\vcenter{\hbox{\BBBOMEpic{B_2}{B_3}{B_1}{B_5}{B_6}{B_4}}} +
\vcenter{\hbox{\BBBOMEpic{B_2}{B_1}{B_3}{B_5}{B_4}{B_6}}}
-\Ps_{13}\Bigg(
\vcenter{\hbox{\BBBOMEpic{B_2}{B_3}{B_1}{B_5}{B_4}{B_6}}} +
\vcenter{\hbox{\BBBOMEpic{B_2}{B_1}{B_3}{B_5}{B_6}{B_4}}}
\Bigg) \\
&\vec q_{63}\colon&&
\vcenter{\hbox{\BBBOMEpic{B_3}{B_1}{B_2}{B_6}{B_4}{B_5}}} +
\vcenter{\hbox{\BBBOMEpic{B_3}{B_2}{B_1}{B_6}{B_5}{B_4}}}
-\Ps_{12}\Bigg(
\vcenter{\hbox{\BBBOMEpic{B_3}{B_1}{B_2}{B_6}{B_5}{B_4}}} +
\vcenter{\hbox{\BBBOMEpic{B_3}{B_2}{B_1}{B_6}{B_4}{B_5}}}
\Bigg) \\
&\vec q_{51}\colon&
\Ps_{23}\Ps_{13}\Bigg(&
\vcenter{\hbox{\BBBOMEpic{B_1}{B_2}{B_3}{B_5}{B_6}{B_4}}} +
\vcenter{\hbox{\BBBOMEpic{B_1}{B_3}{B_2}{B_5}{B_4}{B_6}}}
\Bigg)
-\Ps_{12}\Bigg(
\vcenter{\hbox{\BBBOMEpic{B_1}{B_2}{B_3}{B_5}{B_4}{B_6}}} +
\vcenter{\hbox{\BBBOMEpic{B_1}{B_3}{B_2}{B_5}{B_6}{B_4}}}
\Bigg) \\
&\vec q_{62}\colon&
\Ps_{23}\Ps_{13}\Bigg(&
\vcenter{\hbox{\BBBOMEpic{B_2}{B_3}{B_1}{B_6}{B_4}{B_5}}} +
\vcenter{\hbox{\BBBOMEpic{B_2}{B_1}{B_3}{B_6}{B_5}{B_4}}}
\Bigg)
-\Ps_{23}\Bigg(
\vcenter{\hbox{\BBBOMEpic{B_2}{B_3}{B_1}{B_6}{B_5}{B_4}}} +
\vcenter{\hbox{\BBBOMEpic{B_2}{B_1}{B_3}{B_6}{B_4}{B_5}}}
\Bigg) \\
&\vec q_{43}\colon&
\Ps_{23}\Ps_{13}\Bigg(&
\vcenter{\hbox{\BBBOMEpic{B_3}{B_1}{B_2}{B_4}{B_5}{B_6}}} +
\vcenter{\hbox{\BBBOMEpic{B_3}{B_2}{B_1}{B_4}{B_6}{B_5}}}
\Bigg)
-\Ps_{13}\Bigg(
\vcenter{\hbox{\BBBOMEpic{B_3}{B_1}{B_2}{B_4}{B_6}{B_5}}} +
\vcenter{\hbox{\BBBOMEpic{B_3}{B_2}{B_1}{B_4}{B_5}{B_6}}}
\Bigg) \\
&\vec q_{61}\colon&
\Ps_{23}\Ps_{12}\Bigg(&
\vcenter{\hbox{\BBBOMEpic{B_1}{B_2}{B_3}{B_6}{B_4}{B_5}}} +
\vcenter{\hbox{\BBBOMEpic{B_1}{B_3}{B_2}{B_6}{B_5}{B_4}}}
\Bigg)
-\Ps_{13}\Bigg(
\vcenter{\hbox{\BBBOMEpic{B_1}{B_2}{B_3}{B_6}{B_5}{B_4}}} +
\vcenter{\hbox{\BBBOMEpic{B_1}{B_3}{B_2}{B_6}{B_4}{B_5}}}
\Bigg) \\
&\vec q_{42}\colon&
\Ps_{23}\Ps_{12}\Bigg(&
\vcenter{\hbox{\BBBOMEpic{B_2}{B_3}{B_1}{B_4}{B_5}{B_6}}} +
\vcenter{\hbox{\BBBOMEpic{B_2}{B_1}{B_3}{B_4}{B_6}{B_5}}}
\Bigg)
-\Ps_{12}\Bigg(
\vcenter{\hbox{\BBBOMEpic{B_2}{B_3}{B_1}{B_4}{B_6}{B_5}}} +
\vcenter{\hbox{\BBBOMEpic{B_2}{B_1}{B_3}{B_4}{B_5}{B_6}}}
\Bigg) \\
&\vec q_{53}\colon\qquad &
\Ps_{23}\Ps_{12}\Bigg(&
\vcenter{\hbox{\BBBOMEpic{B_3}{B_1}{B_2}{B_5}{B_6}{B_4}}} +
\vcenter{\hbox{\BBBOMEpic{B_3}{B_2}{B_1}{B_5}{B_4}{B_6}}}
\Bigg)
-\Ps_{23}\Bigg(
\vcenter{\hbox{\BBBOMEpic{B_3}{B_1}{B_2}{B_5}{B_4}{B_6}}} +
\vcenter{\hbox{\BBBOMEpic{B_3}{B_2}{B_1}{B_5}{B_6}{B_4}}}
\Bigg)
\end{alignat*}
\vspace{-1.5\baselineskip}
\caption{
Feynman diagrams contributing to the one-meson exchange three-body potential for \(B_1B_2B_3\to B_4B_5B_6\).
\label{fig:omefeyn}
}
\end{figure*}
\balancecolsandclearpage
}

\section{Two-meson exchange} \label{sec:tme}

For the two-meson exchange diagram of \fig{fig:3BF} we need in addition to the Lagrangian in \eq{eq:LBBM} the well-known \(\mathcal O(q^2)\) meson-baryon Lagrangian \ct*{Krause1990}.
We use the version given in \ct{Oller2006} and display here only the terms relevant for our purpose:
\begin{align*} \label{eq:MBMBLagr}
\mathcal L ={}
& b_D\langle\bar B\{\chi_+,B\}\rangle
+ b_F\langle\bar B[\chi_+,B]\rangle
+ b_0\langle\bar BB\rangle\,\langle\chi_+\rangle \\
&+ b_1\langle\bar B[u^\mu,[u_\mu,B]]\rangle
+ b_2\langle\bar B\{u^\mu,\{u_\mu,B\}\}\rangle \\
&+ b_3\langle\bar B\{u^\mu,[u_\mu,B]\}\rangle
+ b_4\langle\bar BB\rangle\,\langle u^\mu u_\mu\rangle \\
&+\mathrm i d_1\langle\bar B\{[u^\mu,u^\nu],\sigma_{\mu\nu}B\}\rangle
+\mathrm i d_2\langle\bar B[[u^\mu,u^\nu],\sigma_{\mu\nu}B]\rangle \\
&+\mathrm i d_3\langle\bar Bu^\mu\rangle\langle u^\nu\sigma_{\mu\nu}B\rangle \,, \numberthis
\end{align*}
with \(u_\mu = -\frac1{f_0}\partial_\mu\phi+\mathcal O(\phi^3)\) and \(\chi_+ = 2\chi-\frac1{4f_0^2}\{\phi,\{\phi,\chi\}\}\linebreak[0]+\mathcal O(\phi^4)\), where
\begin{equation}
 \chi = \begin{pmatrix} m_\pi^2 & 0 & 0 \\ 0 & m_\pi^2 & 0 \\ 0 & 0 & 2m_K^2-m_\pi^2 \end{pmatrix} \,.
\end{equation}
Note that the terms proportional to \(b_D,b_F,b_0\) break explicitly SU(3) flavor symmetry, through different meson masses \(m_K\neq m_\pi\).
Rewriting the Lagrangian in the particle basis as in the previous sections, one obtains
\begin{align*}
 \mathcal L &= -\sum_{c^f=b_D,b_F,b_0} \frac{c^f}{4f_0^2} \sum_{i,j,k,l=1}^8 N^f_{\phi_k\substack{i\\j}\phi_l} (\bar B_i B_j) \phi_k\phi_l\\
 &+ \sum_{c^f=b_1,b_2,b_3,b_4} \frac{c^f}{f_0^2} \sum_{i,j,k,l=1}^8 N^f_{\phi_k\substack{i\\j}\phi_l} (\bar B_i B_j) \partial_\mu \phi_k \partial^\mu \phi_l\\
 &+ \sum_{c^f=d_1,d_2,d_3} \frac{\mathrm i\,c^f}{f_0^2} \sum_{i,j,k,l=1}^8 N^f_{\phi_k\substack{i\\j}\phi_l} (\bar B_i \sigma_{\mu\nu}B_j) \partial^\mu \phi_k \partial^\nu \phi_l \,. \numberthis
\end{align*}

\begin{figure}
\vspace{.3\baselineskip}
\centering
\begin{overpic}[scale=.6]{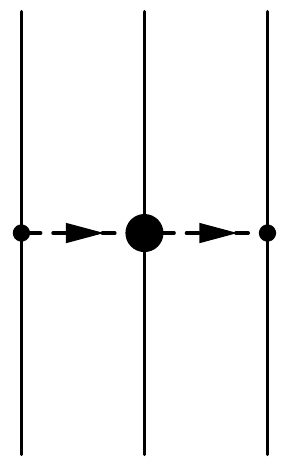}
\put(2,101){$l$}\put(27,101){$m$}\put(55,101){$n$}
\put(2,-8){$i$}\put(27,-8){$j$}\put(55,-8){$k$}
\put(0,-21){$A$}\put(25,-21){$B$}\put(53,-21){$C$}
\put(14,57){$\phi_1$} \put(40,57){$\phi_2$}
\end{overpic}
\vspace{.8\baselineskip}
\caption{
Generic two-meson exchange diagram.
\label{fig:tme-gen}
}
\end{figure}

Let us now consider the generic two-meson exchange diagram depicted in \fig{fig:tme-gen}.
It includes the baryons \(i,j,k\) in the initial state, the baryons \(l,m,n\) in the final state, and two virtual mesons \(\phi_1\) and \(\phi_2\) are exchanged.
The indices \(A,B,C\) label the spin spaces related to the baryon bilinears and they are defined by the three initial baryons.
The momentum transfers carried by the virtual mesons are \(\vec q_{li} = \vec p_l-\vec p_i\) and \(\vec q_{nk} = \vec p_n-\vec p_k\).
One obtains the following transition amplitude from the generic two-meson exchange diagram
\begin{align*}
V ={} & -\frac{1}{4f_0^4} \frac{\vec\sigma_A\cdot\vec q_{li}\ \vec\sigma_C\cdot\vec q_{nk}}{(\vec q_{li}^{\,2}+m_{\phi_1}^2)(\vec q_{nk}^{\,2}+m_{\phi_2}^2)} \\
&\times \Big(N'_1 + N'_2\,\vec q_{li}\cdot\vec q_{nk} +N'_3\,\mathrm i\,(\vec q_{li}\times\vec q_{nk})\cdot\vec\sigma_B\Big) \,, \numberthis
\end{align*}
with the combinations of parameters
\begin{align*}
N'_1 ={}& N_{B_lB_i\bar\phi_1}N_{B_nB_k\phi_2} \\ &\times \sum_{c^f=b_D,b_F,b_0}\frac{c^f}{4} ( N^f_{\phi_1\substack{m\\j}\bar\phi_2} + N^f_{\bar\phi_2\substack{m\\j}\phi_1}) \,, \\
N'_2 ={}& -N_{B_lB_i\bar\phi_1}N_{B_nB_k\phi_2} \\ &\times \sum_{c^f=b_1,b_2,b_3,b_4}c^f ( N^f_{\phi_1\substack{m\\j}\bar\phi_2} + N^f_{\bar\phi_2\substack{m\\j}\phi_1}) \,, \\
N'_3 ={}& N_{B_lB_i\bar\phi_1}N_{B_nB_k\phi_2} \\ &\times \sum_{c^f=d_1,d_2,d_3}c^f ( N^f_{\phi_1\substack{m\\j}\bar\phi_2} - N^f_{\bar\phi_2\substack{m\\j}\phi_1}) \,. \numberthis
\end{align*}

The complete three-body potential for a transition \(B_1B_2B_3\rightarrow B_4B_5B_6\) is finally obtained by summing up the contributions of the 18 Feynman diagrams in \fig{fig:tmefeyn} and by summing over all possible exchanged mesons.
Obviously, additional (negative) spin-exchange operators need to be applied if the baryon lines are not in the configuration 1-4, 2-5 and 3-6, as illustrated in \fig{fig:tmefeyn}.

\afterpage{
\begin{figure}
\centering
\begin{align*}
&
\vcenter{\hbox{\BBBTMEpic{B_1}{B_2}{B_3}{B_4}{B_5}{B_6}}} +
\vcenter{\hbox{\BBBTMEpic{B_2}{B_3}{B_1}{B_5}{B_6}{B_4}}} +
\vcenter{\hbox{\BBBTMEpic{B_3}{B_1}{B_2}{B_6}{B_4}{B_5}}}
\\
+\Ps_{23}\Ps_{13}\Bigg(&
\vcenter{\hbox{\BBBTMEpic{B_1}{B_2}{B_3}{B_5}{B_6}{B_4}}} +
\vcenter{\hbox{\BBBTMEpic{B_2}{B_3}{B_1}{B_6}{B_4}{B_5}}} +
\vcenter{\hbox{\BBBTMEpic{B_3}{B_1}{B_2}{B_4}{B_5}{B_6}}}
\Bigg)\\
+\Ps_{23}\Ps_{12}\Bigg(&
\vcenter{\hbox{\BBBTMEpic{B_1}{B_2}{B_3}{B_6}{B_4}{B_5}}} +
\vcenter{\hbox{\BBBTMEpic{B_2}{B_3}{B_1}{B_4}{B_5}{B_6}}} +
\vcenter{\hbox{\BBBTMEpic{B_3}{B_1}{B_2}{B_5}{B_6}{B_4}}}
\Bigg)\\
-\Ps_{23}\Bigg(&
\vcenter{\hbox{\BBBTMEpic{B_1}{B_2}{B_3}{B_4}{B_6}{B_5}}} +
\vcenter{\hbox{\BBBTMEpic{B_2}{B_3}{B_1}{B_6}{B_5}{B_4}}} +
\vcenter{\hbox{\BBBTMEpic{B_3}{B_1}{B_2}{B_5}{B_4}{B_6}}}
\Bigg)\\
-\Ps_{13}\Bigg(&
\vcenter{\hbox{\BBBTMEpic{B_1}{B_2}{B_3}{B_6}{B_5}{B_4}}} +
\vcenter{\hbox{\BBBTMEpic{B_2}{B_3}{B_1}{B_5}{B_4}{B_6}}} +
\vcenter{\hbox{\BBBTMEpic{B_3}{B_1}{B_2}{B_4}{B_6}{B_5}}}
\Bigg)\\
-\Ps_{12}\Bigg(&
\vcenter{\hbox{\BBBTMEpic{B_1}{B_2}{B_3}{B_5}{B_4}{B_6}}} +
\vcenter{\hbox{\BBBTMEpic{B_2}{B_3}{B_1}{B_4}{B_6}{B_5}}} +
\vcenter{\hbox{\BBBTMEpic{B_3}{B_1}{B_2}{B_6}{B_5}{B_4}}}
\Bigg)
\end{align*}
\vspace{-1.5\baselineskip}
\caption{
Feynman diagrams contributing to the two-meson exchange three-body potential for \(B_1B_2B_3\to B_4B_5B_6\).
\label{fig:tmefeyn}
}
\end{figure}
}

\section{\texorpdfstring{\boldmath \(NNN\)}{N-N-N} and \texorpdfstring{\boldmath \(\Lambda NN\)}{Lambda-N-N} three-baryon potentials} \label{sec:potex}

In order to give a concrete example we present in this section the explicit expressions for the \(\Lambda NN\) three-body interaction in spin-, isospin- and momentum-space.
Moreover, the leading order chiral three-nucleon interaction is rederived, and consistency with the conventional expression is shown.
The potentials are calculated in particle basis (as shown in the previous sections) and afterwards reexpressed with isospin operators.

By adding up all 36 contributions (coming from \eqs{eq:potD} and \eq*{eq:BBBfullpot}), one obtains the following form of the three-nucleon contact potential
\begin{align*}
V^{NNN}_\mathrm{ct} ={}& -\frac38E\,\big[ \\
\phantom{+}(\, 3\, & \mathbbm1 - \vec\sigma_1\cdot\vec\sigma_2 - \vec\sigma_1\cdot\vec\sigma_3 - \vec\sigma_2\cdot\vec\sigma_3 )\ \mathbbm1 \\
+\, ( -& \mathbbm1 - \vec\sigma_1\cdot\vec\sigma_2 + \vec\sigma_1\cdot\vec\sigma_3 + \vec\sigma_2\cdot\vec\sigma_3 )\ \vec\tau_1\cdot\vec\tau_2 \\
+\, ( -& \mathbbm1 + \vec\sigma_1\cdot\vec\sigma_2 - \vec\sigma_1\cdot\vec\sigma_3 + \vec\sigma_2\cdot\vec\sigma_3 )\ \vec\tau_1\cdot\vec\tau_3 \\
+\, ( -& \mathbbm1 + \vec\sigma_1\cdot\vec\sigma_2 + \vec\sigma_1\cdot\vec\sigma_3 - \vec\sigma_2\cdot\vec\sigma_3 )\ \vec\tau_2\cdot\vec\tau_3 \\
  - & \vec\sigma_1\times\vec\sigma_2\cdot\vec\sigma_3\ \vec\tau_1\times\vec\tau_2\cdot\vec\tau_3\, \big] \,, \numberthis
\end{align*}
with the low-energy constant \(E=2(\lc_4-\lc_9) = -c_{\overline{35}}/3\) and \(\vec\sigma\), \(\vec\tau\) denote the usual Pauli matrices in spin and isospin space.
This is exactly the three-nucleon contact potential of \ct{Epelbaum2002} in its antisymmetrized form:
\begin{equation}
V^{NNN}_\mathrm{ct} = \frac12 E\, \mathcal A\, \sum_{j\neq k}\vec\tau_j\cdot\vec\tau_k \,,
\end{equation}
where \(\mathcal A\) denotes the three-body antisymmetrization operator, \(\mathcal A = (\mathbbm1-\mathcal P_{12}) (\mathbbm1-\mathcal P_{13}-\mathcal P_{23})\).
Each two-particle exchange operator
\(\mathcal P_{ij} = \Ps_{ij}\Pt_{ij}\Pp_{ij}\)
is the product of an exchange operator in spin space \(\Ps_{ij}=\frac12(\mathbbm1+\vec\sigma_i\cdot\vec\sigma_j)\), in isospin space \(\Pt_{ij}=\frac12(\mathbbm1+\vec\tau_i\cdot\vec\tau_j)\) and in momentum space \(\Pp_{ij}\).
Note that the leading-order 3\(N\) contact potential is momentum-independent, and therefore \(\Pp_{ij}\) has no effect.
We remind that in our calculation the generalized Pauli principle is automatically built in by performing all Wick contractions.

For the \(\Lambda NN\) contact interaction we obtain the following expression:
\begin{align*}
V^{\Lambda NN}_\mathrm{ct} ={}
& \phantom{{}+{}} \lc'_1\ (\mathbbm1 - \vec\sigma_2\cdot\vec\sigma_3 ) ( 3 + \vec\tau_2\cdot\vec\tau_3 ) \\
& + \lc'_2\ \vec\sigma_1\cdot(\vec\sigma_2+\vec\sigma_3)\,(\mathbbm1 - \vec\tau_2\cdot\vec\tau_3) \\
& + \lc'_3\ (3 + \vec\sigma_2\cdot\vec\sigma_3 ) ( \mathbbm1 - \vec\tau_2\cdot\vec\tau_3 ) \,, \numberthis
\end{align*}
where the primed constants are given by
\begin{align*}
\lc^\prime_1  ={}& -\frac{1}{48} (2 \lc_2 -13 \lc_3 +21 \lc_4 -13 \lc_5 +24 \lc_7 \\& +54 \lc_8 -21 \lc_9 + 54 \lc_{10} -30 \lc_{12} -15 \lc_{13} \\& -15 \lc_{14} -48 \lc_{16} +18 \lc_{17} -18 \lc_{18} ) \,, \\ 
\lc^\prime_2  ={}& -\frac{1}{24} (8 \lc_2 -5 \lc_3 -3 \lc_4 -5 \lc_5 +12 \lc_7 -18 \lc_8 \\& +15 \lc_9 -18 \lc_{10} -3 \lc_{13} -3 \lc_{14} +6 \lc_{17} -6 \lc_{18} ) \,, \\ 
\lc^\prime_3  ={}& -\frac{1}{48} (10 \lc_2 -13 \lc_3 +21 \lc_4 -13 \lc_5 +24 \lc_7 \\& -18 \lc_8 +3 \lc_9 -18 \lc_{10} +18 \lc_{12} -15 \lc_{13} \\& -15 \lc_{14} -6 \lc_{17} +6 \lc_{18} )
\,. \numberthis
\end{align*}
The constants \(\lc_1\ldots\lc_{18}\) originate from the minimal contact Lagrangian in \eq{eq:minct}.
Note that the constant \(\lc'_1\) belongs exclusively to the transition with total isospin \(I=1\), whereas the constants \(\lc'_2\) and \(\lc'_3\) appear for total isospin \(I=0\).
Interestingly, none of these three constants can be substituted by the constant \(E\) of the purely nucleonic sector.
Thus, the strength of the \(\Lambda NN\) three-body contact interaction is not related to the one for \(NNN\) via SU(3) symmetry.

The one-pion exchange three-nucleon potential reads (in antisymmetrized form)
\begin{align*} \label{eq:NNNope}
V^{NNN}_\mathrm{OPE} = \qquad\qquad\quad\quad ( & X_{123}^{456} + X_{231}^{564} + X_{312}^{645} ) \\
+ \Ps_{23}\Pt_{23}\Ps_{13}\Pt_{13} ( &  X_{123}^{564} + X_{231}^{645} + X_{312}^{456} ) \\
+ \Ps_{23}\Pt_{23}\Ps_{12}\Pt_{12} ( &  X_{123}^{645} + X_{231}^{456} + X_{312}^{564} ) \,, \numberthis
\end{align*}
where we have defined the abbreviation%
\footnote{We have used the symbol \(d'\) instead of the conventional \(D\) in order to avoid confusion with the axial vector constant in \eq{eq:LBBM}.}
\begin{align*}
& X_{ijk}^{lmn} =  \\ & \quad
-\frac{g_A}{16f_0^2} d'
\frac{\vec\sigma_i\cdot\vec q_{li}}{\vec q_{li}^{\,2}+m_\pi^2}
\bigg[  (\vec\tau_j-\vec\tau_k)\cdot\vec\tau_i\ (\vec\sigma_j-\vec\sigma_k)\cdot\vec q_{li} \\ & \qquad\qquad\qquad
+ (\vec\tau_j\times\vec\tau_k)\cdot\vec\tau_i\ (\vec\sigma_j\times\vec\sigma_k)\cdot\vec q_{li}  \bigg] \,, \numberthis
\end{align*}
with \(g_A=D+F\) and \(d'=4 (\ld_1 - \ld_3 + \ld_8 - \ld_{10})\).
Each term in \eq{eq:NNNope} corresponds to a complete row in \fig{fig:omefeyn}.
We have verified that this result is equal to the antisymmetrization of the expression given in \ct{Epelbaum2002}, \pagebreak[0]
\begin{equation}
V^{NNN}_\mathrm{OPE} =-\frac{g_A}{8f_\pi^2} d' \mathcal{A} \sum_{i\neq j\neq k}\frac{\vec\sigma_j\cdot\vec q_j}{\vec q_j^{\,2}+m_\pi^2} \vec\tau_i\cdot\vec\tau_j\ \vec\sigma_i\cdot\vec q_j \,,
\end{equation}
inserting the momentum transfers \(\vec q_1 = \vec q_{41} = \vec p_4-\vec p_1\), \(\vec q_2 = \vec q_{52} = \vec p_5-\vec p_2\), \(\vec q_3 = \vec q_{63} = \vec p_6-\vec p_3\).
In this case the momentum part of each two-body exchange operator, \(\Pp_{ij}\), exchanges also the momenta in the final state.%
\footnote{For example, \(\Pp_{23}\) leads to the replacements \(q_{41},q_{52},q_{63}\to q_{41},q_{62},q_{53}\) and \(\Pp_{12}\Pp_{13}\) to \(q_{41},q_{52},q_{63}\to q_{61},q_{42},q_{53}\).}

Let us continue with the \(\Lambda NN\) one-pion exchange three-body potentials.
Many diagrams are absent due to the vanishing of the \(\Lambda\Lambda\pi\)-vertex (by isospin symmetry).
We find the following result for the \(\Lambda NN\) three-body interaction mediated by one-pion exchange:
\begin{widetext}
\begin{align*} \label{eq:LNNope}
V^{\Lambda NN}_\mathrm{OPE} ={} -\frac{g_A}{2f_0^2} \,\bigg( \qquad\qquad
&\frac{\vec\sigma_2\cdot\vec q_{52}}{\vec q_{52}^{\,2}+m_\pi^2} \vec\tau_2\cdot\vec\tau_3
\Big[  (\ld'_1\vec\sigma_1+\ld'_2\vec\sigma_3)\cdot\vec q_{52} \Big] \\
+&\frac{\vec\sigma_3\cdot\vec q_{63}}{\vec q_{63}^{\,2}+m_\pi^2} \vec\tau_2\cdot\vec\tau_3
\Big[  (\ld'_1\vec\sigma_1+\ld'_2\vec\sigma_2)\cdot\vec q_{63} \Big] \\
+\Ps_{23}\Pt_{23} \Ps_{13}&\frac{\vec\sigma_2\cdot\vec q_{62}}{\vec q_{62}^{\,2}+m_\pi^2} \vec\tau_2\cdot\vec\tau_3
\Big[  -\frac{\ld'_1+\ld'_2}2 (\vec\sigma_1+\vec\sigma_3)\cdot\vec q_{62} + \frac{\ld'_1-\ld'_2}2\,\mathrm i\,(\vec\sigma_3\times\vec\sigma_1)\cdot\vec q_{62}  \Big] \\
+\Ps_{23}\Pt_{23} \Ps_{12}&\frac{\vec\sigma_3\cdot\vec q_{53}}{\vec q_{53}^{\,2}+m_\pi^2} \vec\tau_2\cdot\vec\tau_3
\Big[  -\frac{\ld'_1+\ld'_2}2 (\vec\sigma_1+\vec\sigma_2)\cdot\vec q_{53} - \frac{\ld'_1-\ld'_2}2\,\mathrm i\,(\vec\sigma_1\times\vec\sigma_2)\cdot\vec q_{53}  \Big]
\bigg)\,, \numberthis
\end{align*}
\end{widetext}
where we have defined the two linear combinations of constants
\begin{align*}
\ld'_1  ={}& \frac{1}{6} (-3 \ld_1  +\ld_2 +\ld_3 +5 \ld_4 +9 \ld_5 \\& +\ld_6 -6 \ld_8 +\ld_{11} +2 \ld_{12} -3 \ld_{13} ) \,, \\
\ld'_2  ={}& \frac{1}{6} (\ld_1 +\ld_2 -3 \ld_3 +\ld_4 +9 \ld_5 +5 \ld_6 \\& -6 \ld_{10} -\ld_{11} -2 \ld_{12} +3 \ld_{13} ) \,. \numberthis
\end{align*}
The four lines in \eq{eq:LNNope} correspond to the four rows in \fig{fig:omefeyn} that have no \(\Lambda\) hyperon at the baryon-baryon-meson vertex, \ie the diagrams involving the momentum transfers \(\vec q_{52}\), \(\vec q_{63}\), \(\vec q_{62}\), \(\vec q_{53}\).

Finally, we obtain for the three-nucleon interaction mediated by two-pion exchange
\begin{align*} \label{eq:tmeN}
V^{NNN}_\mathrm{TPE} = \qquad\qquad\qquad
& \left( Y_{123}^{456} + Y_{231}^{564} + Y_{312}^{645} \right) \\
+ \Ps_{23}\Pt_{23} \Ps_{13}\Pt_{13}
& \left( Y_{123}^{564} + Y_{231}^{645} + Y_{312}^{456} \right) \\
+ \Ps_{23}\Pt_{23} \Ps_{12}\Pt_{12}
& \left( Y_{123}^{645} + Y_{231}^{456} + Y_{312}^{564} \right) \\
- \Ps_{23}\Pt_{23}
& \left( Y_{123}^{465} + Y_{231}^{654} + Y_{312}^{546} \right) \\
- \Ps_{13}\Pt_{13}
& \left( Y_{123}^{654} + Y_{231}^{546} + Y_{312}^{465} \right) \\
- \Ps_{12}\Pt_{12}
& \left( Y_{123}^{546} + Y_{231}^{465} + Y_{312}^{654} \right)
\,, \numberthis
\end{align*}
where the eighteen terms follow the ordering displayed in \fig{fig:tmefeyn} and we have introduced the abbreviation
\begin{align*}
Y_{ijk}^{lmn} ={}&
\frac{g_A^2}{4f_\pi^4} \frac{\vec\sigma_i\cdot\vec q_{li}\ \vec\sigma_k\cdot\vec q_{nk}}{ (\vec q_{li}^{\,2}+m_\pi^2)(\vec q_{nk}^{\,2}+m_\pi^2)}\\ & \quad
\times\Big[ \vec\tau_i\cdot\vec\tau_k (-4 c_1 m_\pi^2+2c_3\vec q_{li}\cdot\vec q_{nk}) \\ & \qquad
\quad+ c_4 \vec\tau_j\cdot(\vec\tau_i\times\vec\tau_k)\ \vec\sigma_j\cdot(\vec q_{li}\times\vec q_{nk})
\Big] \,, \numberthis
\end{align*}
with the constants (see also \cts{Frink2004,Mai2009})
\begin{align*}
 c_1 &= \frac{1}{2} (2 b_0+b_D+b_F)\,,\\
 c_3 &= b_1+b_2+b_3+2 b_4\,, \\
c_4 &= 4 (d_1+d_2)\,. \numberthis
\end{align*}

Again, the result in \eq{eq:tmeN} is equal to the antisymmetrization of the expression given in \ct{Epelbaum2002}:
\begin{equation}
V^{NNN}_\mathrm{TPE} = \frac{g_A^2}{8f_\pi^2} \mathcal{A} \sum_{i\neq j\neq k} \frac{\vec\sigma_i\cdot\vec q_i\ \vec\sigma_j\cdot\vec q_j}{ (\vec q_i^{\,2}+m_\pi^2)(\vec q_j^{\,2}+m_\pi^2)} F^{\alpha\beta}_{ijk}\tau^\alpha_i\tau^\beta_j \,,
\end{equation}
with
\begin{align*}
& F^{\alpha\beta}_{ijk} =
 \frac{\delta^{\alpha\beta}}{f_\pi^2} (-4 c_1 m_\pi^2+2c_3\vec q_i\cdot\vec q_j) \\
 &\qquad\quad +\sum_\gamma \frac{c_4}{f_\pi^2}\epsilon^{\alpha\beta\gamma}\tau_k^\gamma\ \vec\sigma_k\cdot(\vec q_i\times\vec q_j) \,. \numberthis
\end{align*}

The \(\Lambda NN\) three-body interaction generated by two-pion exchange takes the form
\begin{align*}
V^{\Lambda NN}_\mathrm{TPE} ={} & \\ &
\frac{g_A^2}{3f_0^4}
\frac{\vec\sigma_3\cdot\vec q_{63}\ \vec\sigma_2\cdot\vec q_{52}}{(\vec q_{63}^{\,2}+m_{\pi}^2)(\vec q_{52}^{\,2}+m_{\pi}^2)} \vec\tau_2\cdot\vec\tau_3 \\ &
\ \times\Big(     -(3 b_0 + b_D) m_\pi^2      +      (2 b_2 + 3 b_4)      \,\vec q_{63}\cdot\vec q_{52}\Big) \\
- \Ps_{23}\Pt	_{23} & \frac{g_A^2}{3f_0^4}
\frac{\vec\sigma_3\cdot\vec q_{53}\ \vec\sigma_2\cdot\vec q_{62}}{(\vec q_{53}^{\,2}+m_{\pi}^2)(\vec q_{62}^{\,2}+m_{\pi}^2)} \vec\tau_2\cdot\vec\tau_3 \\ &
\ \times\Big(     -(3 b_0 + b_D) m_\pi^2      +      (2 b_2 + 3 b_4)      \,\vec q_{53}\cdot\vec q_{62}\Big) \,. \numberthis
\end{align*}
Note that only those two diagrams in \fig{fig:tmefeyn} contribute, where the (final and initial) \(\Lambda\) hyperon are associated to the central baryon line.
All other diagrams are simply zero due to the vanishing of the \(\Lambda \Lambda \pi\) vertex.

\section{Summary and Outlook} \label{sec:sum}

In this work we have derived the leading contributions to the three-baryon interaction from SU(3) chiral effective field theory.
First, we have established the minimal non-relativistic Lagrangian for contact terms of six octet-baryons, leading to 18 constants.
Using this foundation, general SU(3) relations among the three-baryon channels with strangeness \(0\) and \(-1\) have been derived.
Furthermore, the four-baryon contact Lagrangian with one Goldstone boson has been given in its minimal form, in which it involves 14 constants.
The irreducible three-body potentials have been constructed at next-to-next-to-leading order in the chiral power counting based on the effective chiral Lagrangians.
Contributions arise from contact terms, from one-meson exchange and from two-meson exchange diagrams.
The three-body potential for the \(\Lambda NN\) interaction has been presented as an explicit example.

The large number of unknown low-energy constants is related to the variety of three-baryon multiplets, with strangeness ranging from \(0\) to \(-6\).
For selected processes only a small subset of these constants contributes as has been exemplified for the \(\Lambda NN\) three-body interaction.
Estimates of the predominant low-energy constants can be made by using decuplet-baryon saturation.
An example for that is the \(\Sigma^*(1385)\) excitation in case of the \(\Lambda NN\) two-pion exchange three-body interaction.
Due to the small decuplet-octet mass splitting, such effects are promoted to next-to-leading order in the chiral power counting, in analogy to the role played by the \(\Delta\) resonance in the nucleonic sector \ct*{Epelbaum2009}.
Work along this direction is in progress \ct*{PetschauerPrep}.
We anticipate that the chiral potentials derived in this work will shed light on the role of three-baryon forces in hypernuclei.
In particular, their application in studies of light hypernuclei will be very instructive because such systems can be treated within reliable few-body techniques \ct*{Nogga2014a,Wirth2014}.
Furthermore, one expects that the present investigations can help paving the way for more systematic studies on the role of three-baryon interactions in hyperonic neutron star matter.

\begin{acknowledgments}
We thank A.~Nogga for useful discussions.
This work is supported in part by DFG and NSFC through funds provided to the Sino-German CRC~110 ``Symmetries and the Emergence of Structure in QCD''.

\end{acknowledgments}

\end{document}